\documentclass[a4paper,11pt]{article}
\usepackage[margin=1in]{geometry}
\usepackage{amssymb, amsmath, amsthm, mathrsfs}
\usepackage{cite}
\usepackage{lineno}
\usepackage{tabls}
\usepackage{multirow}
\usepackage{pifont}
\usepackage[ruled]{algorithm2e}
\usepackage{graphicx}
\usepackage{url}
\usepackage{color}
\definecolor{myblue}{rgb}{0,0,1}
\definecolor{myred}{rgb}{1,0,0}

\def\qed{\hfill $\Box$\vspace{0.3cm}}
\def\pf{\noindent{\bf Proof. }}

\newtheorem{lemma}{Lemma}[section]

\newtheorem{theorem}{Theorem}
\newtheorem{remark}{Remark}[section]

\newtheorem{definition}{Definition}

\begin{document}

\title{\LARGE\bf Transmission Failure Analysis of Multi-Protection Routing in Data Center Networks with Heterogeneous Edge-Core Servers}
\author{\small
Xiao-Yan Li$^{1}$
\hspace{.05in} Wanling Lin$^{1}$
\hspace{.05in} Jou-Ming Chang$^{2}$ 
\hspace{.05in} Xiaohua Jia$^{3}$ 
\\ \\
{\small $^1$ College of Mathematics and Computer Science, Fuzhou University, Fuzhou 350108, China}\\
{\small $^2$ Institute of Information and Decision Sciences,}\\
{\small National Taipei University of Business, Taipei 10051, Taiwan}\\
{\small $^3$ Department of Computer Science, City University of Hong Kong, Hong Kong}
}
\date{}
\maketitle

\begin{abstract}
\baselineskip=12pt

The recently proposed RCube network is a cube-based server-centric data center network (DCN), including two types of heterogeneous servers, called core servers and edge servers. Remarkably, it takes the latter as backup servers to deal with server failures and thus achieve high availability. This paper first points out that RCube is suitable as a candidate topology of DCNs for edge computing. Three transmission types are among core and edge servers based on the demand for applications' computation and instant response. We then employ protection routing to analyze the transmission failure of RCube DCNs. Unlike traditional protection routing, which only tolerates a single link or node failure, we use the multi-protection routing scheme to improve fault-tolerance capability. To configure a protection routing in a network, according to Tapolcai's suggestion, we need to construct two completely independent spanning trees (CISTs), which are edge-disjoint and inner-vertex-disjoint spanning trees. It is well-known that the problem of determining whether there exists a dual-CIST (i.e., two CISTs) in a network is NP-complete. A logic graph of RCube, denoted by $L$-$RCube(n,m,k)$, is a network with a recursive structure. Each basic building element consists of $n$ core servers and $m$ edge servers, where the order $k$ is the number of recursions applied in the structure. In this paper, we provide algorithms to construct $\min\{n,\lfloor(n+m)/2\rfloor\}$ CISTs in $L$-$RCube(n,m,k)$ for $n+m\geqslant 4$ and $n>1$. From a combination of the multiple CISTs, we can configure the desired multi-protection routing. In our simulation, we configure up to 10 protection routings for RCube DCNs. As far as we know, in past research, there were at most three protection routings developed in other network structures. Finally, we summarize some crucial analysis viewpoints about the transmission efficiency of DCNs with heterogeneous edge-core servers from the simulation results.

\end{abstract}
\setcounter{page}{1}
\baselineskip=14pt

\section{Introduction}
\label{sec:intro}

Edge Computing aims to move the computation away from cloud data centers towards the edge of the network so that specific cloud services can attain the more efficient quality, particularly a reduction in the data transmission, processing rate, and network traffic. Whether the adoption of centralized cloud or distributed edge servers, \emph{data center network} (DCN) is fundamental and plays a pivotal role in a data center, as it interconnects all of the computation and storage resources together. Recently, many novel networking structures for large-scale DCNs have been proposed, e.g., see~\cite{Chen-2016,Xia-2016} for surveys. According to the difference of inherent computing tasks assigned to switches or servers, DCNs can be divided into two categories, namely, \emph{switch-centric DCNs} and \emph{server-centric DCNs}. Because of the different service requirements and constraints, each type of DCNs has its own application scope. For data transmission in switch-centric networks, switches set up a transmission path and invoke a route to relay packets for communications between servers, while servers are only responsible for packet forwarding. The classical FatTree~\cite{Al-Fares-2008}, VL2~\cite{Greenberg-2009}, PortLand~\cite{Mysore-2009} and Jupiter rising~\cite{Singh-2015} belong to this category. In contrast, for data transmission in server-centric networks, servers play as the source, destination, and relay nodes for determining a dedicated transmission path, and all switches act merely as crossbars that transfer packets according to the selected path. The well-known server-centric DCNs include DCell~\cite{DCell-2008}, BCube\cite{Guo-2009}, FiConn~\cite{Li-2011}, Dpillar~\cite{Liao-2012}, HCN and BCN~\cite{Guo-2013}, SWcube~\cite{Li-2015}, and HSDC~\cite{Zhang-2019}.

The significant advantage of server-centric DCNs is that it can reduce hardware costs, consume less energy, and has a higher degree of programmability. This paper focuses on discussing a novel server-centric DCNs called \emph{RCube}, which attracted our attention because it is more flexible in the design of configuration~\cite{Li-2018}. An RCube network can be constructed through a recursive structure, and it possesses many favorable properties. The primitive concept of designing RCube was based on server failures' backup requirements; however, it naturally generalized another type of DCNs called BCube (defined later in Section~\ref{sec:BCube}). Generally, as the network's scale increases, inevitable component failures (including switches and servers) will become more frequent. Singh et al.~\cite{Singh-2015} pointed out that servers are more prone to failure than switches. The most effective solution is to add redundant servers to the DCN as a backup for overcoming the problem of server failure. According to this requirement, Li and Yang~\cite{Li-2018} recently proposed a solution for RCube DCNs. Meanwhile, they have pointed out that high availability becomes an even crucial challenge for designing a server-centric network with backup servers. The main reason is that each server is responsible for forwarding packets, and putting some servers in standby mode may cut off the transmission path, thus further affecting the throughput of the entire system. The favorable features of RCube DCNs presented in~\cite{Li-2018} include the following:

\begin{itemize}
\item It has abundant multiple parallel paths, and its diameter is linearly proportional to the network order. The above factors make the communication between end servers in an RCube lower latency.
\item Compared to BCube, it is flexible to make trade-off among power consumption and aggregate throughput and deliver a similar performance of critical metrics, such as average path length and path distribution.
\item It can configure edge servers to the active/passive redundancy mode, and thus the availability of the entire system can be significantly enhanced.
\end{itemize}
To the best of our knowledge, RCube is the only server-centric network supporting the last feature.

\begin{figure}[ht]
\centering
 \includegraphics[width=3.3in]{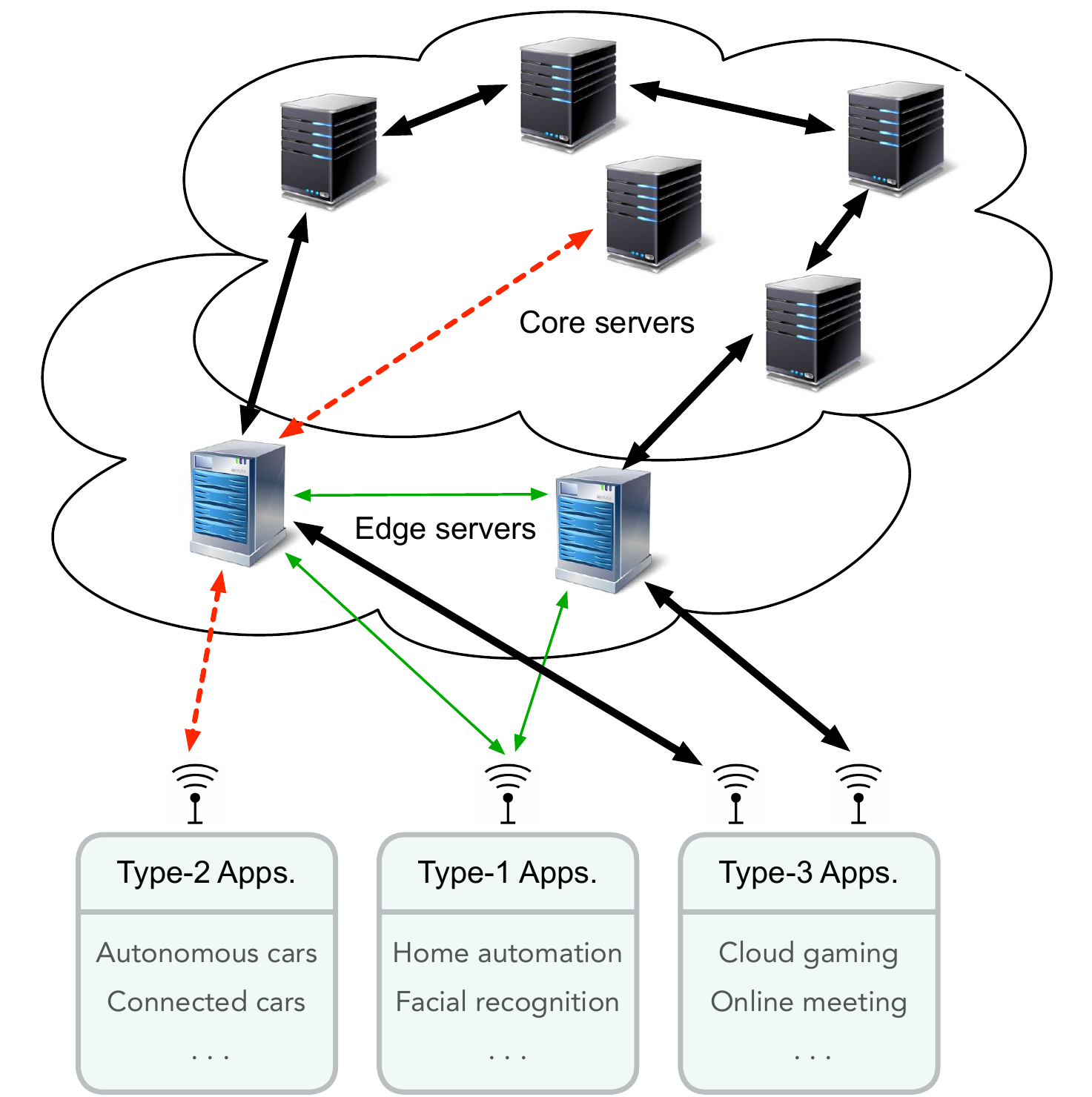}
 \caption{Three types of transmission in a DCN with heterogeneous edge-core servers.}
 \label{fig:edge-core}
\end{figure}

For improving data availability in a cloud environment, effective replica management is also another solution to resolve server failures~\cite{Gill-2016,Li-2020}. This paper intends to solve fault-tolerant transmission in RCube DCNs without putting servers in standby mode. Compared with the servers planned for backup in the original, we treat them as another heterogeneous server called \emph{edge servers}, which can deal with fewer computing services. On the contrary, we regard non-backup servers as \emph{core servers}, which have more power on computation and can handle more complex tasks. Just as cloud computing operates on ``big data'' while edge computing operates on ``instant data'', we take advantage of this notion to RCube DCNs. Note that edge and core servers are complementary components of a whole computing system. Based on this idea, the ratio of data transmission between different servers will vary with applications. For instance, specific instant applications, such as home automation systems~\cite{Chakraborty-2017} and facial recognition algorithms~\cite{Qian-2019}, need to carry out computation and transmission only in local edge servers. Some real-time applications, such as short-term autonomous and connected cars~\cite{Yang-2019}, need to perform calculation, synchronization, navigation, and sending messages between core servers and edge servers. Most online applications, such as cloud gaming (or gaming on demand)~\cite{Anand-2017} and online video meetings (or web conferencing), require vast amounts of visual computation and communication among multiple core servers and edge servers. Therefore, we consider three types of transmission among core servers and edge servers, as illustrated in Fig.~\ref{fig:edge-core}.

According to the above three types of transmission, we will employ \emph{protection routing} to carry out the transmission failure analysis of fault-tolerant routing for RCube DCNs. Kwong et al.~\cite{Kwong-2011} introduced protection routing for integrating route discovery and route maintenance mechanisms in mobile ad-hoc networks (MANETs). Notably, a protection routing uses a centralized and reactive routing protocol fitting for intra-domain IP networks. Since a protection routing exploits the multi-paths technique for fault-tolerance, it guarantees a loop-free alternate path for packet forwarding when a single link or node failure occurs. Pai et al.~\cite{Pai-2020-JPDC,Pai-2020-IS} recently demonstrated that protection routing is also suitable for relatively large (static) network topologies with scalability, such as networks with recursive structure. Tapolcai~\cite{Tapolcai-2013} showed that a graph or network possessing two \emph{completely independent spanning trees} (CISTs) is sufficient to configure a protection routing. A collection of CISTs is a set of unrooted spanning trees in a graph such that paths connecting every pair of vertices in these trees are edge-disjoint and inner-vertex-disjoint. However, Hasunuma~\cite{Hasunuma-2002} early proved that the problem of determining whether there exists $k$ CISTs in a graph is NP-complete, even for $k=2$ (i.e., a dual-CIST). For more researches on constructing a dual-CIST or multiple CISTs on interconnection networks, please refer to~\cite{Mane-2018,Pai-2019-TON,Pai-2020-JPDC,Pai-2020-IS,Pai-2021,Pai-2019-TCS} and references quoted therein. In particular, the works in~\cite{Pai-2019-TON,Pai-2020-JPDC,Pai-2020-IS,Pai-2021} discussed in-depth the configuration of protection routing. In contrast, to the best of our knowledge, only a few studies of CISTs are related to DCNs. For server-centric DCNs, recent results showed the existence of a dual-CIST on DCell~\cite{Qin-2019} and multiple CISTs, respectively, on HSDC~\cite{Qin-2020} and DCNs based on augmented cubes~\cite{Chen-2021}. The formal definitions and more discussions of CISTs and protection routing will be given in Sections~\ref{sec:CIST} and \ref{sec:protection-route}, respectively.

In this paper, an RCube is represented by its logic graph $L$-$RCube(n,m,k)$ for ease of description.  The parameters $n$ and $m$ denote the numbers of core servers and edge servers in a basic building element, and the order $k$ is the number of recursions applied in RCube. Then, we propose two algorithms for constructing multiple CISTs in RCube. One is for the base graph $L$-$RCube(n,m,1)$, and the other is for high-dimensional $L$-$RCube(n,m,k)$ with $k\geqslant 2$. In each of the cases, the number of CISTs we constructed is $\min\{n,\lfloor(n+m)/2\rfloor\}$ for $n+m\geqslant 4$ and $n>1$, and the running time of the algorithm is $\mathcal{O}(n^k(n+m))$. Using the combination of these multiple CISTs, we can configure the desired multi-protection routing for our simulation to increase fault-tolerance capability. Technically, the use of $k$ CISTs can configure $k\choose 2$ different protection routings. So far, the simulation of multi-protection routing has just employed in a relatively smaller interconnection network, and it only configures at most three protection routings (e.g., see~\cite{Pai-2020-JPDC,Pai-2020-IS}). However, the existing work on multi-protection routing for server-centric networks has not yet progressed. This paper is the first time to design multi-protection routing on a server-centric network, and our simulation can reach up to 10 protection routings.

The rest of this paper is organized as follows. Section~\ref{sec:prelim} formally gives the definitions of BCube, RCube, protection routing, and CISTs. It also describes Tapolcai's method for configuring a network's protection routing and introduces some useful properties related to CISTs. Section 3 presents an algorithm to construct multiple CISTs for base graph $L$-$RCube(n,m,1)$. Section~\ref{sec:main} extends the construction of multiple CISTs in high-dimensional $L$-$BCCC(n,m,k)$ for $k\geqslant 2$ by recursion. Section~\ref{sec:App} provides some simulations to analyze transmission failures for three types of transmission in the protection routing of RCube. Meanwhile, it also affords comprehensive performance evaluations for protection routing in RCube DCNs. Finally, we offer some concluding remarks in the last section.

\section{Preliminaries}
\label{sec:prelim}

In this section, we introduce some definitions, terminologies, and notations. We first provide Table~\ref{tbl:notation} that describes some of the important notations used in this paper. For convenience, the terms ``networks'' and ``graphs'', and ``vertices'' and ``servers'' are often used interchangeably.

\begin{table}[ht]
\caption{Summary of Notation}
\begin{center}
\begin{tabular}{|l|l|} \hline
\multicolumn{1}{|c|}{Symbol} & \multicolumn{1}{|c|}{Description} \\ \hline
$\mathbb{N}^+$ & The set of all positive integers. \\
$[n]$ & The set of integers $\{1,2,\ldots,n\}$. \\
$[n,m]$ & The set of integers $\{n,n+1,\ldots,m\}$ where $n<m$. \\
$V(G)$ & The vertex set of a graph $G$. \\
$E(G)$ & The edge set of a graph $G$. \\
$(u,v)$ & The edge with two ends $u$ and $v$. \\
$N_G(v)$ & The open neighborhood of the vertex $v$ in $G$. \\
$\text{deg}_{G}(v)$ & The degree of the vertex $v$ in $G$ (or simply $d(v)$). \\
$\text{diam}(G)$ & The diameter of the graph $G$. \\
$G-f$ & The graph obtained from $G$ by removing a failed \\
& component $f$. \\
$G\cong H$ & Two graphs $G$ and $H$ are isomorphic. \\
$R_d$ & A routing with a destination vertex $d$. \\
{\sc pnh}($u$) & The primary next-hop of a vertex $u$ in a routing. \\
{\sc pl}($u$) & The primary link of a vertex $u$ in a routing. \\
{\sc snh}($u$) & The second next-hop of a vertex $u$ in a routing. \\
$K_{n}$ & The complete graph with $n$ vertices. \\
\hline
\end{tabular}
\end{center}
\label{tbl:notation}
\end{table}%

\subsection{Overview of BCube DCNs Structure}
\label{sec:BCube}

\begin{figure}[ht]
\centering
 \includegraphics[width=3.5in]{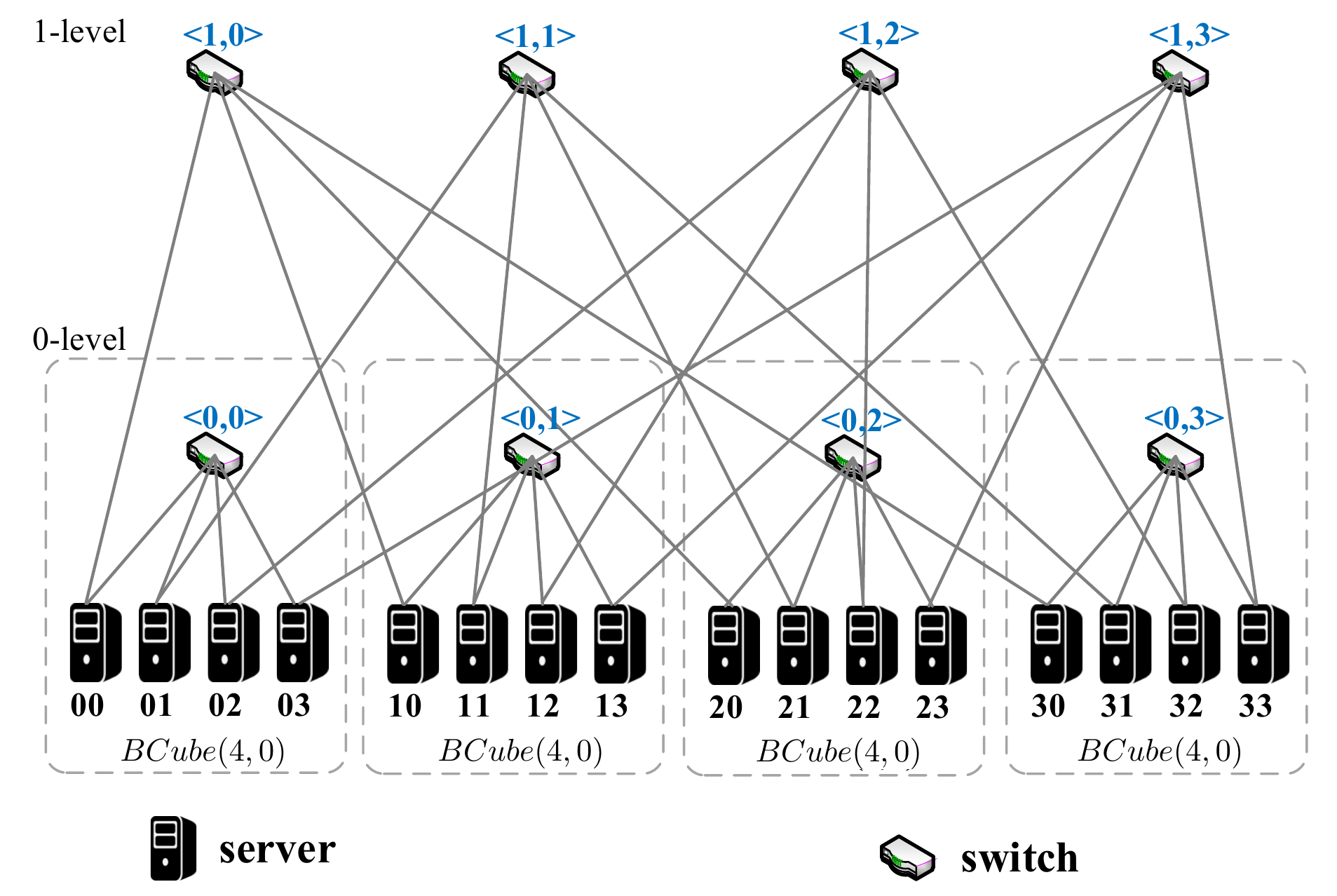}
 \caption{A BCube data center network $BCube(4,1)$ constructed by dual-port servers and 4-port switches. The rectangle boxes drawn by dashed lines represent $BCube(4,0)$'s.}
 \label{fig:BCube}
\end{figure}

We first introduce some background of the preliminary prototype of RCube. About a decade ago, Guo et al.~\cite{Guo-2009} proposed a DCN named \emph{BCube}, which possesses high aggregate throughput, low diameter, and abundant parallel paths. The structure of a BCube can be described by two parameters $n$ and $k$ and denoted by $BCube(n,k)$. It consists of $n^{k+1}$ $(k+1)$-port servers and $(k+1)n^k$ $n$-port switches, and the latter are evenly divided into $k+1$ levels. In each level of $BCube(n,k)$, there are $n^k$ switches, and each server connects one switch at every level through one of its ports based on the designed principle. Meanwhile, we can also define BCube in a recursive fashion. The basic building element of BCube is $BCube(n,0)$, which is composed of $n$ servers connected by an $n$-port switch. In general, $BCube(n,k)$ is constructed from $n$ $BCube(n,k-1)$'s with $n^k$ additional switches. For more details on the connection rule between switches and servers in a BCube, the reader can refer to~\cite{Guo-2009}. Fig.~\ref{fig:BCube} depicts an example of $BCube(4,1)$.

\subsection{Overview of Logic RCube DCNs Topology}
\label{sec:RCube}

An \emph{RCube} DCN consists of a massive number of servers and switches equipped with multiple line cards and network interface controller (NIC) ports, respectively. Similar to BCube, an RCube can be constructed by a recursive structure. For an RCube, its \emph{order} (or \emph{dimension}) is the number of recursions applied in the construction. We use the notation $RCube(n,m,k)$ to denote a RCube with order $k$ $(k\in \mathbb{N}^+)$, where $n$ is the number of core servers and $m$ is the number of edge servers, and these servers are connected by an $(n+m)$-port switch in a basic building element $RCube(n,m,0)$. Obviously, $BCube(n,k)$ is a special case of $RCube(n,m,k)$ when $m=0$. For $RCube(n,m,k)$, each server connects to $(k+1)$ switches and each switch connects to $(n+m)$ servers, and thus there are $(n+m)n^k$ servers and $(k+1)n^k$ switches in total. The following is the formal definition of RCube DCNs.

\begin{definition}\label{def:RCube}{\rm(See~\cite{Li-2018})
For $k\geqslant 1$, $RCube(n,m,k)$ is composed of $n$ $RCube(n,m,k-1)$'s connected with $n^k$ $(n+m)$-port switches according to the following rules:
\begin{enumerate}
\item[(1)] The addresses of servers are denoted as $a_{k}a_{k-1}\cdots a_{0}$, where $a_0\in [0,n+m-1]$ and $a_{i}\in [0,n-1]$ for $1\leqslant i\leqslant k$. In particular, a server is a core server if $a_0<n$; otherwise, it is an edge server.
\item[(2)] The addresses of switches are represented as $s_{k}s_{k-1}\cdots s_{0}$, where $s_{k}\in [0,k]$ and $s_{i}\in [0,n-1]$ for $0\leqslant i\leqslant k-1$.
\item[(3)] Each core server is equipped with $k+1$ NIC ports numbered by $0$ to $k$, which are used to connect to a switch in the corresponding level under the condition
$a_ka_{k-1}\cdots a_{s_k+1} a_{s_k-1}\cdots a_0=s_{k-1} s_{k-2}\cdots s_0$.
\item[(4)] Edge servers are the same as core servers except that they are connected to switches under the condition $a_k a_{k-1}\cdots a_1=s_{k-1} s_{k-2}\cdots s_0$.
\end{enumerate}
}
\end{definition}

\begin{figure}[ht]
\centering
\includegraphics[width=3.3in]{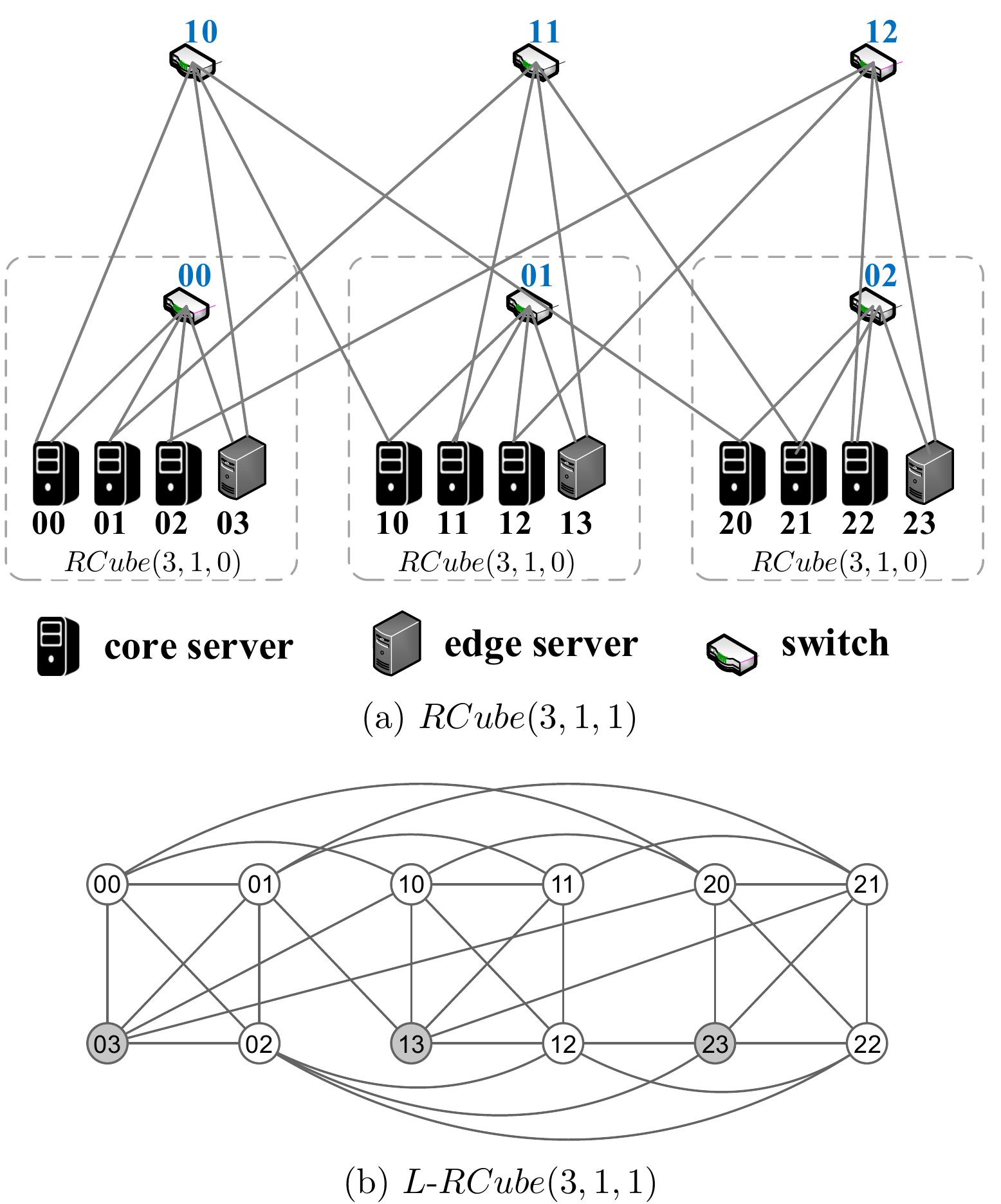}
\caption{(a) An RCube data center network $RCube(3,1,1)$ constructed by dual-port servers and 4-port switches. The rectangle boxes drawn by dashed lines represent $RCube(3,1,0)$'s; (b) The logic graph $L$-$RCube(3,1,1)$, where the white vertices represent core servers and the gray vertices represent edge servers.}
\label{fig:RCube}
\end{figure}

Fig.~\ref{fig:RCube}(a) shows an example of $RCube(3,1,1)$, which is composed of three $RCube(3,1,0)$'s and three 4-port switches. There are three core servers and one edge server in each $RCube(3,1,0)$. In $RCube(3,1,1)$, server 20 is a core server, and by Definition~\ref{def:RCube}(3), it connects to switch 02 through port at level 0 (because $a_1=s_0=2$) and connects to switch 10 through port at level 1 (because $a_0=s_0=0$). Also, server 03 is an edge server, and by Definition~\ref{def:RCube}(4), it connects to switches 00 and 10 (because $a_1=s_0=0$).

%

A \emph{complete graph} is a graph in which every two vertices are joined by an edge. A complete graph with $n$ vertices is denoted by $K_n$. Since all switches have no processing power and act simply as crossbars for transferring packets from one side to another, they can be regarded as transparent devices. Thus, $RCube(n,m,0)$ is abstracted as a complete graph $K_{n+m}$. Moreover, from the adjacency of servers connecting by switches in Definition~\ref{def:RCube}, the logic graph of the network $RCube(n,m,k)$, denoted as $L$-$RCube(n,m,k)$, can be defined as follows.

\begin{definition}\label{definition2}{\rm
The graph $L$-$RCube(n,m,k)$ is composed of $n^k$ $K_{n+m}$'s and it has the vertex set
\[
\{v_kv_{k-1}\cdots v_0\colon\,v_0\in[0,n+m-1]\ \text{and}\ v_i\in[0,n-1], i\in[k]\},
\]
and two vertices $u=u_{k}u_{k-1}\cdots u_0$ and $v=v_{k}v_{k-1}\cdots v_0$ are adjacent if and only if one of the following conditions holds:
\begin{enumerate}
\item[(1)] $u_ku_{k-1}\cdots u_1 = v_kv_{k-1}\cdots v_1$ and $u_0\ne v_0$;
\item[(2)] $u_{i-1}\cdots u_0 = v_{i-1}\cdots v_0$, $u_{i+1} = v_{i+1}$, and $u_i\ne v_i$ for $i\in[k]$ when $u_0<n$;
\item[(3)] $u_{i}u_{i-1}\cdots u_1 = v_{i-1}v_{i-2}\cdots v_0$ and $u_{i+1} = v_{i+1}$ for $i\in[k]$ when $n\leqslant u_0<n+m$.
\end{enumerate}
}
\end{definition}

Fig.~\ref{fig:RCube}(b) depicts $L$-$RCube(3,1,1)$. For example, for any two vertices $u,v\in\{00,01,02,03\}$, since only the rightmost digits of them are different, by Definition~\ref{definition2}(1), $u$ and $v$ are adjacent. Thus $\{00,01,02,03\}$ forms a $K_4$.
For any two vertices $u,v\in\{01,11,21\}$, since only the digits at position $i=1$ are different and $u_0=1<3$, by Definition~\ref{definition2}(2), $u$ and $v$ are adjacent. Thus $\{01,11,21\}$ forms a $K_3$. Also, for vertices $u=03$ and $v\in\{10,20\}$, since $u_1=0=v_0$ and $3=u_0<4$, by Definition~\ref{definition2}(3), $u$ and $v$ are adjacent. Accordingly, the logic graph is non-regular, where each of vertices $00,03,11,13,22$ and $23$ has degree 5, and all other vertices have degree 6.

\subsection{Completely Independent Spanning Trees}
\label{sec:CIST}

A \emph{spanning tree} $T$ in a graph $G$ is an acyclic connected subgraph of $G$ such that $V(T)=V(G)$. Let $t\geqslant 2$ be an integer and $T_1,T_2,\ldots,T_t$ be spanning trees of $G$. A vertex is said to be a \emph{leaf} in $T_i$ if it has degree one, and an \emph{inner-vertex} otherwise. Two spanning trees $T_i$ and $T_j$ are \emph{edge-disjoint} if they share no common edge, and \emph{inner-vertex-disjoint} provided the paths joining any two vertices $x,y\in V(G)$ in both trees have no common vertex except for $x$ and $y$. The spanning trees $T_1,T_2,\ldots,T_t$ are called \emph{completely independent spanning trees} (CISTs for short) if they are pairwise edge-disjoint and inner-vertex-disjoint. In particular, if $t=2$, the set $\{T_1,T_2\}$ is called a dual-CIST. The following characterization is important for studying CISTs.

\begin{theorem}\label{thm:Hasunuma} {\rm(See~\cite{Hasunuma-2001})}
A set of spanning trees $T_1,T_2,\ldots,T_t$ are CISTs of a graph $G$ if and only if they are edge-disjoint and for any vertex $v\in V(G)$, at most one spanning tree $T_i$ for $i\in[t]$ contains $v$ as its inner-vertex.
\end{theorem}

\begin{figure}[ht]
\centering
\includegraphics[width=3in]{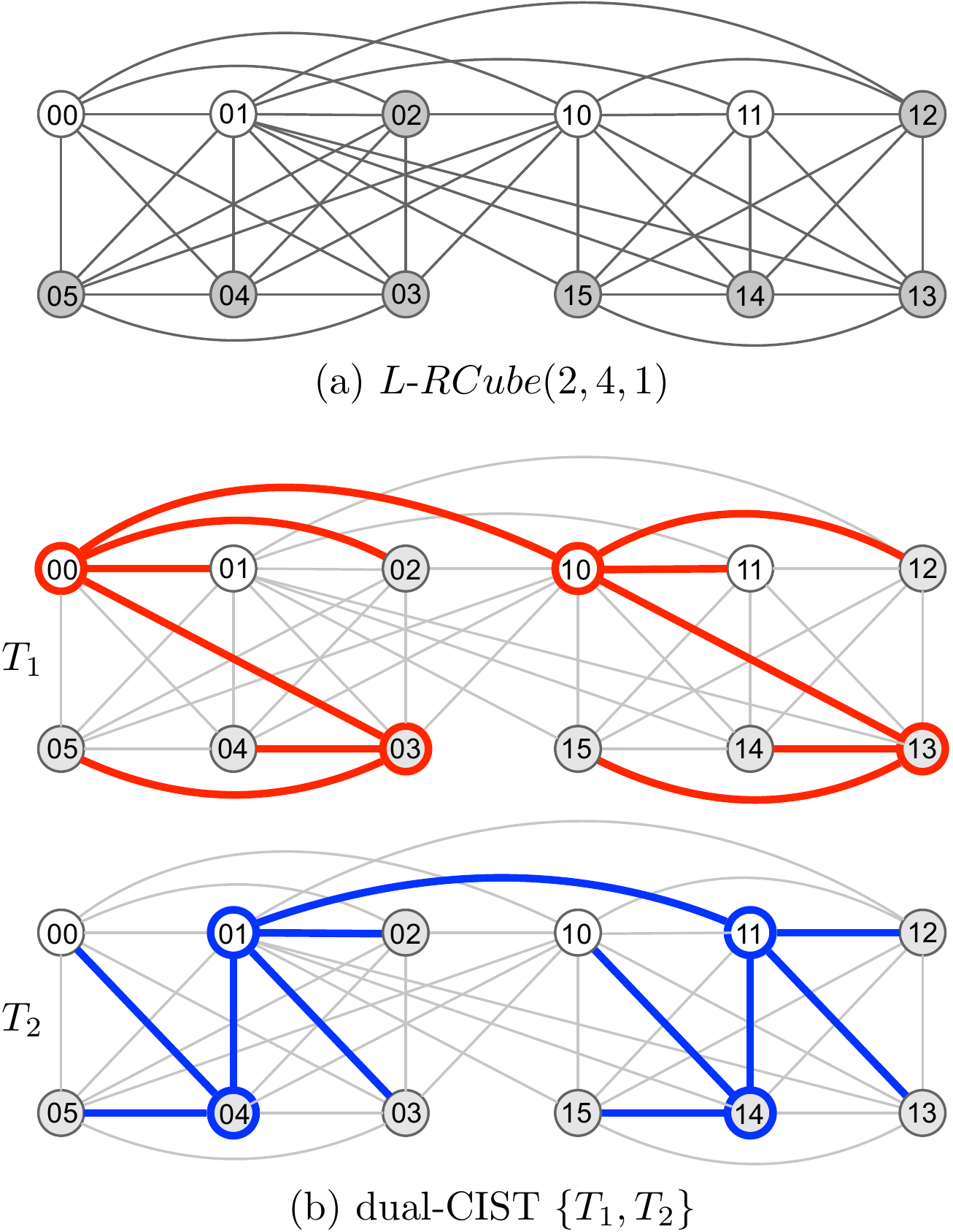}
\caption{(a) The logic graph $L$-$RCube(2,4,1)$, where the white vertices represent core servers and the gray vertices represent edge servers; (b) Two CISTs of $L$-$RCube(2,4,1)$, where bold lines indicate the tree edges and bold line circles indicate inner-vertices.}
\label{fig:2CIST}
\end{figure}

For instance, Fig.~\ref{fig:2CIST}(a) gives another example of logic graph $L$-$RCube(2,4,1)$, and Fig.~\ref{fig:2CIST}(b) shows a dual-CIST $\{T_1,T_2\}$ of $L$-$RCube(2,4,1)$, where $T_1$ and $T_2$ contains $\{00,03,10,13\}$ and $\{01,04,11,14\}$ as sets of inner-vertices, respectively. For BCube and complete graphs, we have known the following results of CISTs.

%
%

\begin{lemma}\label{lem:CIST-BCube}{\rm(See~\cite{Pan-2017})}
There are $\lfloor \frac{n}{2} \rfloor$ CISTs in $L$-$BCube(n,k)$ for $n\geqslant4$ and $k\geqslant1$.
\end{lemma}

\begin{lemma}\label{lem:Kn}{\rm(See~\cite{Pai-2012})}
There are $\lfloor \frac{n}{2} \rfloor$ CISTs in $K_n$ for $n\geqslant4$.
\end{lemma}

In particular, Pai et al.~\cite{Pai-2012} provided a specific construction of $\lfloor n/2\rfloor$ CISTs for $K_{n}$, as shown in Algorithm~\ref{algo1}. Obviously, this algorithm can be run in $\mathcal{O}(n)$ time. Clearly, each spanning tree $T_i$ contains two inner-vertices $v_i$ and $v_{i+t}$. For example, two CISTs of $K_4$ include the following edge sets: $E(T_1)=\{(v_1,v_3),(v_1,v_2),(v_3,v_4)\}$ and $E(T_2)=\{(v_2,v_4),(v_2,v_3),(v_4,v_1)\}$.

\begin{algorithm}
\small
\caption{\small Constructing $\lfloor n/2\rfloor$ CISTs in $K_{n}$}
\label{algo1}
\KwIn{A complete graph $K_{n}$ with $V(K_n)=\{v_1,v_2,\ldots,v_n\}$ for $n\geqslant 4$.}
\KwOut{A set of $t=\lfloor n/2\rfloor$ CISTs $\{T_1,T_2,\dots,T_t\}$.}

\For{$i\leftarrow 1$ {\bf to} $t$}{
	$E(T_i)\leftarrow (v_i,v_{i+t})\cup\{(v_i,v_j)\colon\,j\in[i+1,t+i-1]\}$
	$\cup\;\{(v_{i+t},v_j)\colon\,j\in[1,i-1]\cup[t+i+1,n]\}$\;
}
\Return $\{T_1,T_2,\ldots,T_t\}$.
\end{algorithm}


\subsection{Protection Routing}
\label{sec:protection-route}

Let $G=(V,E)$ be an undirected graph. For $v\in V$, let $N_{G}(v)=\{u\in V\colon\,(u,v)\in E\}$ be the set of neighbors of $v$ in $G$, and let $\text{deg}_G(v)=|N_G(v)|$. For a destination vertex $d\in V$, a routing for traffic destined to $d$ is a directed acyclic graph $R_{d}=(V,E_{d})$ such that the underlying graph of $R_{d}$ is a spanning subgraph of $G$ (i.e., $E_d\subseteq E$) and every vertex $v\in V \backslash\{d\}$ has one outgoing link in $R_{d}$. For notational convenience, we still use $(u,v)$ to stand for the directed link from $u$ to $v$. If $(u,v)\in E_{d}$, the link is called a \emph{primary link} of $u$ and is denoted by {\sc pl}($u$). Also, $v$ is called a \emph{primary next-hop} of $u$ and is denoted by {\sc pnh}($u$). For $u,v\in V$, we say that $u$ is in the \emph{upstream} (resp.\ \emph{downstream}) of $v$ if there is a directed path from $u$ to $v$ (resp.\ from $v$ to $u$) in $R_{d}$. For a failed component $f\in (V\cup E)\backslash \{d\}$, we denote $G-f$ and $R_{d}-f$ as the residual network and routing obtained from $G$ and $R_{d}$ by removing $f$, respectively.

\begin{definition}\label{def:protection}{\rm(See~\cite{Kwong-2011})
A vertex $u\in V\backslash\{d\}$ in a routing $R_{d}$ is \emph{protected} with respect to $d$ if for any single failed component $f=\text{\sc pnh}(u)$ or $f=\text{\sc pl}(u)$,
there is a vertex $w\in N_{G-f}(u)$ such that the following conditions hold:

(i) $w$ is not in the upstream of $u$ in $R_{d}-f$, and

(ii) $w$ and all its downstream vertices (expect $d$) have at least one primary next-hop in $R_{d}-f$.
}
\end{definition}

Note that the vertex $w$ in Definition~\ref{def:protection} is called the \emph{second next-hop} of $u$ and is denoted by {\sc snh}($u$). Condition (i) is to avoid forwarding loops when a component failure occurs, and condition (ii) guarantees that packets are delivered to $d$ through vertex $w$ and its downstream vertices in $R_{d}-f$.

\begin{definition}\label{def:routing}{\rm(See~\cite{Kwong-2011})
A routing $R_{d}$ is a \emph{protection routing} if every vertex $v\in V\backslash\{d\}$ is protected in $R_{d}$. A network $G=(V,E)$ is \emph{protectable} if there exists a protection routing $R_{d}$ for all $d\in V$; otherwise, $G$ is \emph{unprotectable}.
}
\end{definition}

Tapolcai~\cite{Tapolcai-2013} showed that if a dual-CIST of a graph is available, then it is easy to configure a protection routing. Let $\{T_1,T_2\}$ be a dual-CIST of a graph $G=(V,E)$ and $d\in V(G)$ be the destination vertex for traffic. By Theorem~\ref{thm:Hasunuma}, without loss of generality, we assume $d$ is a leaf of $T_2$. Moreover, we can partition $V(G)$ into the following two subsets:
\[
L_1=\{v\in V(G)\colon\,v\ \text{is a leaf only in}\ T_1\}
\]
and
\[
L_2=\{v\in V(G)\colon\,v\ \text{is a leaf in}\ T_2\}.
\]

For $i\in\{1,2\}$, the \emph{skeleton} of $T_i$, denoted by $\hat{T}_i$, is the subtree of $T_i$ that removes all vertices of $L_i$. Since $d$ is a leaf of $T_2$, we let $\ell$ be the unique edge incident with $d$ in $T_2$. Then, $\ell$ connects the two skeletons $\hat{T}_1$ and $\hat{T}_2$. A protection routing $R_d=(V,E_d)$ can be configured by including all links of $\hat{T}_1$, $\hat{T}_2$ and $\ell$ such that the resulting graph takes all links directed to $d$. Algorithm~\ref{Tapolcai} briefly sketches Tapolcai's method.

\begin{algorithm}[h]
\small
\caption{\small Protection Routing}
\label{Tapolcai}
\KwIn{A dual-CIST $\{T_1,T_2\}$ of a network $G=(V,E)$ and the destination vertex $d$.
}
\KwOut{A protection routing $R_d=(V,E_d)$ and {\sc snh}($u$) for all $u\in V(G)\backslash\{d\}$.
}
$L_1\leftarrow\{v\in V(G)\colon\,\text{deg}_{T_1}(v)=1\ \text{and}\ \text{deg}_{T_2}(v)\ne 1\}$\;
$L_2\leftarrow\{v\in V(G)\colon\,\text{deg}_{T_2}(v)=1\}$\;
\For{$i\in\{1,2\}$}{
	$\hat{T}_i\leftarrow T_i-L_i$\;
	\For{$u\in V(G)\backslash\{d\}$}{
		\If{$u\in L_i$}{
			$\text{\sc snh}(u)\leftarrow w$, where $(u,w)\in E(T_i)$;
		}
	}
}
$\ell\leftarrow$ the edge incident with $d$ in $T_2$\;
$E_d\leftarrow E(\hat{T}_1)\!\cup\!E(\hat{T}_2)\!\cup\!\{\ell\}$ such that all links are directed to $d$\;
\Return $R_d=(V,E_d)$ and a table containing {\sc snh}($u$) for all $u\in V(G)\backslash\{d\}$.
\end{algorithm}

Obviously, Algorithm~\ref{Tapolcai} can be run in $\mathcal{O}(|V(G)|)$ time. From the above algorithm, Tapolcai~\cite{Tapolcai-2013} proved the following result.

\begin{theorem}\label{thm:protectable}{\rm(See~\cite{Tapolcai-2013})}
A graph with a dual-CIST is protectable.
\end{theorem}

For example, we consider a dual-CIST of $L$-$RCube(2,4,1)$ as shown in Fig.~\ref{fig:2CIST}(b) and let $d=13$ be the destination vertex for traffic. Then, we have
\begin{eqnarray*}
L_1 & = & \{01,04,11,14\};\\
L_2 & = & \{00,02,03,05,10,12,13,15\}.
\end{eqnarray*}
The sets of links in $\hat{T}_1$ and $\hat{T}_2$ are as follows:
\begin{eqnarray*}
E(\hat{T}_1) & = & \{(00,10),(02,00),(03,00),(05,03),\\
             &   & (10,13),(12,10),(15,13)\}; \\
E(\hat{T}_2) & = & \{(01,11),(04,01),(14,11)\}.
\end{eqnarray*}
Also, we can easily find that $\ell=(11,13)$.

\begin{figure}[ht]
\centering
\includegraphics[width=3.5in]{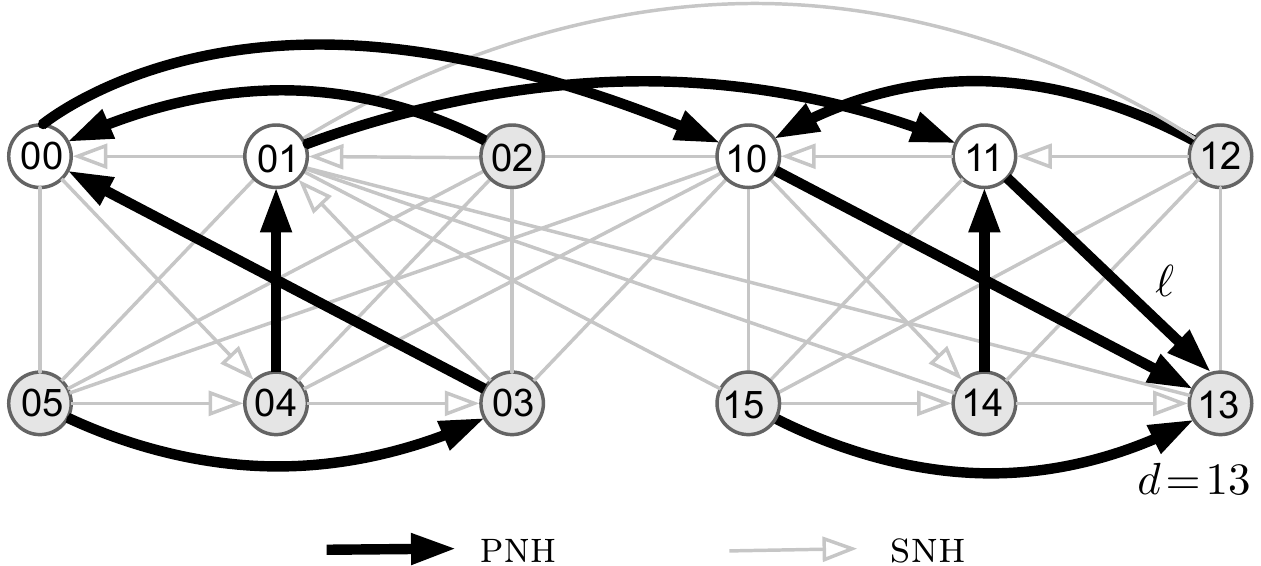}
\caption{A protection routing for traffic destined to the vertex $d=13$ in $L$-$RCube(2,4,1)$, where  a thick link (resp.\ thin link) from $u$ to $v$ indicate $v=\text{\sc pnh}(u)$ (resp.\ $v=\text{\sc snh}(u)$).}
\label{fig:routing}
\end{figure}

Fig.~\ref{fig:routing} shows the protection routing configured by Algorithm~\ref{Tapolcai} for destination vertex $d=13$ in $L$-$RCube(2,4,1)$. For $u\in V(L$-$RCube(2,4,1))\backslash\{d\}$, we can also determine {\sc snh}($u$) by the following rule: for $i\in\{1,2\}$, if $u\in L_i$ is incident with the unique edge $(u,w)\in E(T_i)$, then we have $\text{\sc snh}(u)=w$ in the protection routing. For instance, we have $\text{\sc snh}(03)=01$ because $03\in L_2$ and $(03,01)\in E(T_2)$ (see Fig.~\ref{fig:2CIST}(b)). Similarly, we have $\text{\sc snh}(04)=03$ because $04\in L_1$ and $(04,03)\in E(T_1)$. We now suppose that server 03 is trying to forward packets in the routing, and a failure occurs just right at server $\text{\sc pnh}(03)=00$. In this case, these packets will be immediately forwarded to server $\text{\sc snh}(03)=01$ via the backup route when server 03 detects the event of failure. Similarly, if server 04 is trying to forward packets and it detects the link $\text{\sc pl}(04)=(04,01)$ fails, then these packets will be redirected to server $\text{\sc snh}(04)=03$. We summarize the results of {\sc pnh}($u$), {\sc pl}($u$), and {\sc snh}($u$) for all $u\in V(L$-$RCube(2,4,1))\backslash\{d\}$ in Table~\ref{tbl:2}.

\begin{table}[ht]
\centering
\caption{A summary of primary next-hops, primary links, second next-hops of vertices in the protection routing shown in Fig~\ref{fig:routing}.}
\medskip 
 \begin{tabular}{|c |p{5cm}|c|}
 \hline
  Vertex $u$ & The failure of {\sc pnh}($u$) or {\sc pl}($u$) & {\sc snh}($u$)\\ \hline
  00 & server 10 or link (00,10) fails & 04 \\
  01 & server 11 or link (01,11) fails & 00 \\
  02 & server 00 or link (02,00) fails & 01 \\
  03 & server 00 or link (03,00) fails & 01 \\
  04 & server 01 or link (04,01) fails & 03 \\
  05 & server 03 or link (05,03) fails & 04 \\
  10 & link (10,13) fails & 14 \\
  11 & link (11,13) fails & 10 \\
  12 & server 10 or link (12,10) fails & 11 \\
  14 & server 11 or link (14,11) fails & 13 \\
  15 & link (15,13) fails & 14 \\ \hline
 \end{tabular}
 \label{tbl:2}
\end{table}

\section{Constructing multiple CISTs in $RCube(n,m,1)$}
\label{sec:base}

From the recursive structure, $L$-$RCube(n,m,1)$ can be decomposed into $n$ basic building elements. For convenience, we denote $L$-$RCube(n,m,0)$ as $H_i$ for $i\in[n]$. Clearly, each $H_i$ is isomorphic to a complete graph with $n+m$ vertices (i.e., $H_i\cong K_{n+m}$), and $H_1,H_2,\ldots,H_n$ are vertex-disjoint. Let $V(H_i)=\{v_{i,1},v_{i,2},\ldots,v_{i,n+m}\}$ and map the second indices of the vertices to the servers' addresses in lexicographic order. Based on the previous knowledge, we first points out the existence of multiple CISTs of $RCube(n,m,1)$ for some particular parameters $n$ and $m$.

\begin{itemize}
\item
For $n=1$ and $m\geqslant 3$, since $L$-$RCube(1,m,1)$ is a complete graph with $m+1$ vertices (i.e., $K_{m+1}$), by Lemma~\ref{lem:Kn}, there are $t=\lfloor (m+1)/2 \rfloor$ CISTs.
\item
For $n\geqslant 4$ and $m=0$, we can regard $L$-$RCube(n,0,1)$ as $L$-$BCube(n,1)$. By Lemma~\ref{lem:CIST-BCube}, there are $t=\lfloor n/2 \rfloor$ CISTs.
\item
For $n+m\geqslant 4$, $n\ne 1$, and $m\ne 0$, let $s=\lfloor (n+m)/2\rfloor$ and $t=\min\{n,s\}$. For each $i\in[n]$, since $H_i\cong K_{n+m}$ and $t\leqslant s=\lfloor (n+m)/2\rfloor$, from Algorithm~\ref{algo1}, we can easily construct $t$ CISTs of $H_i$, say $\tilde T_{i,1},\tilde T_{i,2},\dots,\tilde T_{i,t}$, such that each $\tilde T_{i,j}$ contains $v_{i,j}$ and $v_{i,j+s}$ as inner-vertices for $j\in[t]$. Particularly, for each $j\in[t]$, we choose a predestinate index $p$ according to the following rule: $p\in[n]$ if $n\leqslant t$, and $p\in[t]$ otherwise. Customarily, we may choose $p=1$. We call the inner-vertex $v_{p,j}$ of $\tilde T_{p,j}$ the so-called \emph{central vertex}, which corresponds to a core sever by Definition~\ref{def:RCube}. Then, we can construct a spanning tree $T_j$ by connecting $\tilde T_{p,j}$ and $\tilde T_{i,j}$ for all $i\in[n]\backslash\{p\}$ through the edges $(v_{p,j},v_{i,j})$.
\end{itemize}

In the above description, the first two points are apparent. To make the last point clearer, we consider the logic graph $L$-$RCube(2,4,1)$ in Fig.~\ref{fig:2CIST}(a), which consists of two subgraphs $H_1$ and $H_2$, and each is isomorphic to $K_6$. Clearly, $s=3$ and $t=n=2$. By the mapping of servers' addresses, we have
\[
V(H_i)=\{v_{i,j}=(i-1)(j-1)\colon\ j\in[6]\}\ \ \text{for}\ i\in[2].
\]
For $i\in[n]$ and $j\in[t]$, the pairs of inner-vertices are
$\{v_{1,1}=00,v_{1,4}=03\}$ in $\tilde T_{1,1}$, $\{v_{2,1}=10,v_{2,4}=13\}$ in $\tilde T_{2,1}$,
$\{v_{1,2}=01,v_{1,5}=04\}$ in $\tilde T_{1,2}$, and $\{v_{2,2}=11,v_{2,5}=14\}$ in $\tilde T_{2,2}$, respectively. If we choose $p=1$, we have the central vertex $v_{1,1}=00$ for $j=1$ and $v_{1,2}=01$ for $j=2$. As a result, joining the edge $(v_{1,1}=00,v_{2,1}=10)$ between $\tilde T_{1,1}$ and $\tilde T_{2,1}$ and joining the edge $(v_{1,2}=01,v_{2,2}=11)$ between $\tilde T_{1,2}$ and $\tilde T_{2,2}$, we obtain the desired dual-CIST, as shown in Fig.~\ref{fig:2CIST}(b).

From above, we may assume that $n+m\geqslant 4$, and consider three cases as follows: (i) $n=1$; (ii) $m=0$; and (iii) $n\ne 1$ and $m\ne 0$. We now provide detailed steps for constructing these CISTs, as shown in Algorithm~\ref{alg:L-RCube(n,m,1)}.

\newpage
\begin{algorithm}
\small
\caption{\small Constructing multiple CISTs in $RCube(n,m,1)$}
\label{alg:L-RCube(n,m,1)}
\LinesNumbered
\DontPrintSemicolon
\KwIn{A logic graph $L$-$RCube(n,m,1)$ with $n\!+\!m\!\geqslant\!4$, which is decomposed into $n$ subgraphs $H_1,H_2,\ldots,H_n$ and each $H_i (\cong K_{n+m})$ has vertices $v_{i,1},v_{i,2},\ldots,v_{i,n+m}$.}
\KwOut{A set of $t$ CISTs $\{T_1,T_2,\ldots,T_t\}$, where $t=\lfloor(1+m)/2\rfloor$ if $n=1$; and
$t=\min\{n,\lfloor(n+m)/2\rfloor\}$ otherwise.}

$s\leftarrow\lfloor(n+m)/2\rfloor$;\;
\eIf{$n=1$}{
	$t\leftarrow s$;
}
{
	$t\leftarrow \min\{n,s\}$;
}
\For{$i\leftarrow 1\ {\bf to}\ n$}
{
Call {\bf Algorithm~\ref{algo1}} to generate $t$ CISTs $\tilde T_{i,1},\tilde T_{i,2},\dots,\tilde T_{i,t}$ of $H_i$, where $\tilde T_{i,j}$ for $j\in[t]$ contains $v_{i,j}$ and $v_{i,j+s}$ as inner-vertices;
}
\For{$j\leftarrow 1\ {\bf to}\ t$}{
	\uIf{$n=1$}{
		$T_j\leftarrow\tilde T_{1,j}$;
	}
	\uElseIf{$m=0$}{
		$T_j\leftarrow\left(\bigcup\limits_{i=1}^n\tilde{T}_{i,j}\right)
		\cup\left(\bigcup\limits_{i=2}^n\big\{(v_{1,j},v_{i,j}),(v_{1,j+s},v_{i,j+s})\big\}\right)
		\setminus\left(\bigcup\limits_{i=2}^n\big\{(v_{i,j},v_{i,j+s})\big\}\right)$;
	}
	\Else{
		$T_j\leftarrow\left(\bigcup\limits_{i=1}^n\tilde{T}_{i,j}\right)
		\cup\left(\bigcup\limits_{i=2}^n\big\{(v_{1,j},v_{i,j})\big\}\right)$;
	}
}
\Return $\{T_1,T_2,\ldots,T_t\}$.
\end{algorithm}

\begin{theorem}\label{thm:3-main}
For $n+m\geqslant 4$, Algorithm~\ref{alg:L-RCube(n,m,1)} constructs $t$ CISTs of $L$-$RCube(n,m,1)$ in $\mathcal{O}(n(n+m))$ time, where
\[
t=\begin{cases}
\lfloor(1+m)/2\rfloor & \text{if $n=1$}; \\
\min\{n,\lfloor(n+m)/2\rfloor\} & \text{otherwise.}
\end{cases}
\]
\end{theorem}
\pf
Recall that the time complexity of Algorithm~\ref{algo1} is linearly proportional to the number of vertices in the complete graph. Since $L$-$RCube(n,m,1)$ is decomposed into $n$ vertex-disjoint complete subgraphs and each subgraph has $n+m$ vertices, the first for-loop performed in Lines 7-9 requires $\mathcal{O}(n(n+m))$ time. Also, the second for-loop in Lines 10-18 has $t$ iterations, and each iteration has the complexity $\mathcal{O}(n)$ time in the worst case. By the case either $t=\lfloor(1+m)/2\rfloor$ or $t=\min\{n,\lfloor(n+m)/2\rfloor\}$, the loop has $\mathcal{O}(n(n+m))$ running time. Therefore, the time complexity of the algorithm is as desired.

The following is proof of correctness. For the case of $n=1$, this means that $L$-$RCube(n,m,1)$ is a complete graph with $n+m$ vertices. Its correctness directly follows from Lemma~\ref{lem:CIST-BCube} and the construction of Algorithm~\ref{algo1}.

For the case of $m=0$ (i.e., $L$-$BCube(n,1)$), the construction refers to a method in~\cite{Pan-2017} that uses the pair of inner-vertices $v_{1,j}$ and $v_{1,j+s}$ in $\tilde T_{1,j}$ as central vertices to connect $\tilde T_{i,j}$ by two edges for all $i\in[2,n]$. Then, removing edge between vertices $v_{i,j}$ and $v_{i,j+s}$ in $\tilde T_{i,j}$ for all $i\in[2,n]$ to obtain the resulting tree $T_j$ (see Line~14). As mentioned in Remark~\ref{rmk:diameter}, there exist other construction schemes suggested in~\cite{Pan-2017} for building CISTs of $L$-$BCube(n,1)$. Algorithm~\ref{alg:L-RCube(n,m,1)} takes advantage of a construction scheme with a smaller diameter of CISTs.

For $n\ne 1$ and $m\ne 0$, we first construct $\tilde T_{i,1},\tilde T_{i,2},\dots,\tilde T_{i,t}$ as CISTs in the subgraph $H_i$ for $i\in[n]$. Then, adding edge between the central vertex $v_{1,j}$ of $\tilde T_{1,j}$ and the vertex $v_{i,j}$ of $\tilde T_{i,j}$ for all $i\in[n]\backslash\{1\}$ (see Line~16). For each $j\in[t]$, since $\tilde T_{1,j},\tilde T_{2,j},\dots,\tilde T_{n,j}$ are vertex-disjoint, and each $\tilde T_{i,j}$ is a spanning subgraph of $H_i$, the result of connecting all trees $\tilde T_{1,j},\tilde T_{2,j},\dots,\tilde T_{n,j}$ with additional edges is a spanning tree of $L$-$RCube(n,m,1)$. Also, since $\tilde T_{i,1},\tilde T_{i,2},\dots,\tilde T_{i,t}$ are edge-disjoint in $H_i$ for $i\in[n]$, and all connecting edges between trees are different, $T_1,T_2,\ldots,T_t$ are edge-disjoint spanning trees. Moreover, because the two ends of a connecting edge between trees are inner-vertices, it does not change any leaf to become an inner-vertex in the construction. As a result, every vertex of $L$-$RCube(n,m,1)$ can serve as an inner-vertex at most once in the resulting spanning trees. By Theorem~\ref{thm:Hasunuma}, $T_1,T_2,\ldots,T_t$ are CISTs of $L$-$RCube(n,m,1)$.
\qed

\begin{remark}\label{rmk:diameter}{\rm
For $L$-$RCube(n,m,1)$ with $n+m\geqslant 4$, let $T$ be a spanning tree constructed by Algorithm~\ref{alg:L-RCube(n,m,1)}. Then, $T$ has the diameter as follows:
\[
\text{diam}(T)=\begin{cases}
3 & \text{if $n=1$}; \\
5 & \text{if $n=2$ or $m=0$}; \\
6 & \text{otherwise.}
\end{cases}
\]
For the case of $m=0$, there are other construction schemes of CISTs for $L$-$BCube(n,1)$ that were mentioned in~\cite{Pan-2017}. For $n\geqslant 4$, if we change the construction scheme of $T_j$ for $j\in[t]$ from the statement of Line 14 in Algorithm~\ref{alg:L-RCube(n,m,1)} to that of Line 16, the diameter will increase by one. Also, if the change is the following
\[
T_j\leftarrow\left(\bigcup\limits_{i=1}^n\tilde{T}_{i,j}\right)
\cup\left(\bigcup\limits_{i=1}^{n-1}\big\{(v_{i,j},v_{i+1,j})\big\}\right),
\]
then the diameter will increase by two. However, for $n\ne 1$ and $m\ne 0$, we cannot change the construction scheme of $T_j$ from the statement of Line 16 to that of Line 14 because the edge $(v_{1,j+s},v_{i,j+s})$ does not exist between two edge servers when $j+s>n$.
}
\end{remark}

\begin{figure}[t]
\centering
\includegraphics[width=3.2in]{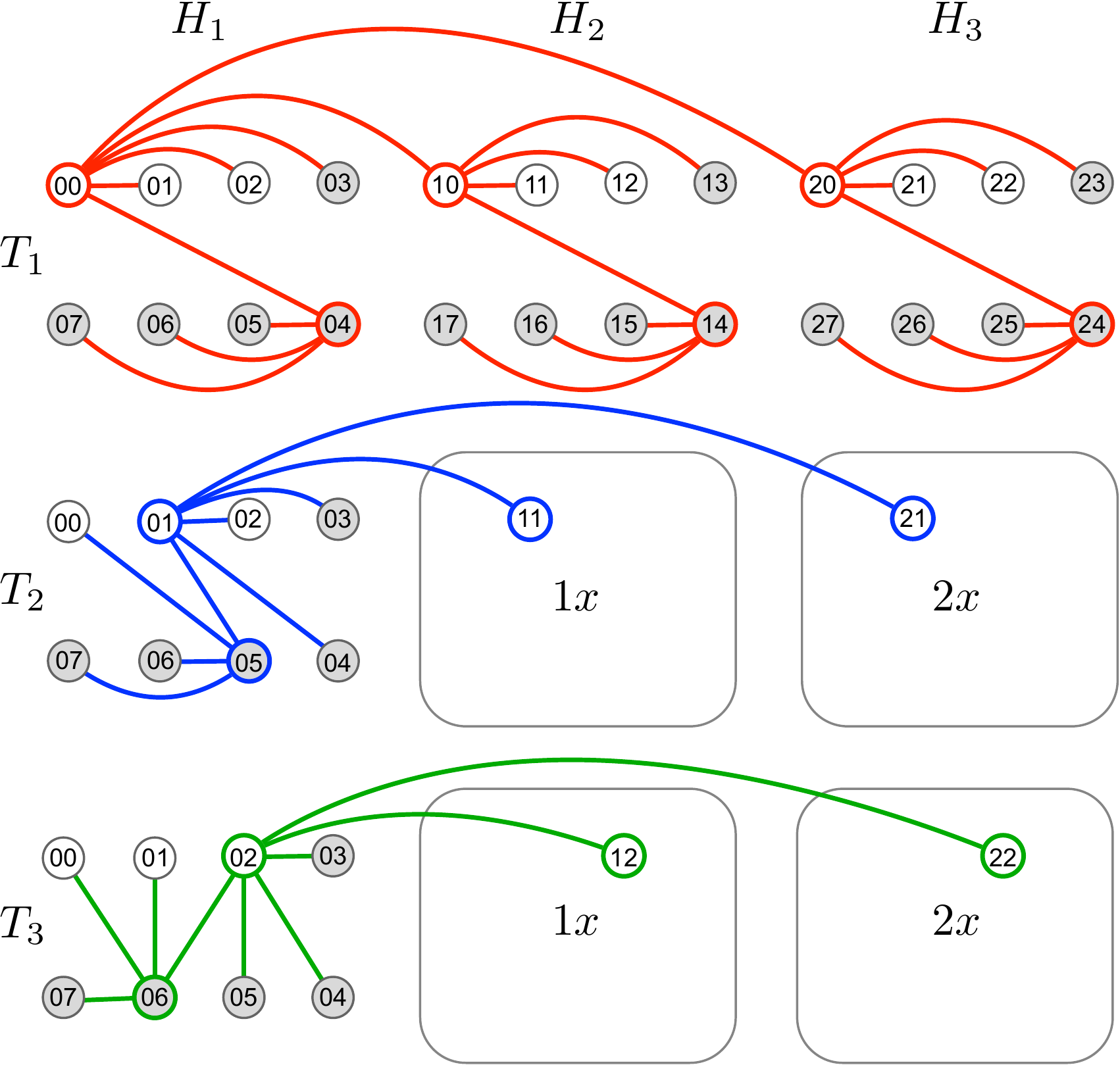}
\caption{Three CISTs $T_1,T_2,T_3$ of $L$-$RCube(3,5,1)$ constructed by Algorithm~\ref{alg:L-RCube(n,m,1)}. In $T_j$ where $j\in\{2,3\}$, a round-corner box with label $1x$ (resp.\ $2x$) represents a subtree isomorphic to $T_{1,j}$ and it contains vertices $1x$ (resp.\ $2x$) for all $x\in[0,7]$.}
\label{3CIST-351}
\end{figure}

Before ending this section, we give another example of CISTs constructed by Algorithm~\ref{alg:L-RCube(n,m,1)}. Fig.~\ref{3CIST-351} shows the three CISTs of $L$-$RCube(3,5,1)$ in a way of the partial view. It is easy to observe that for any $j\in[t]$, the trees $T_{i,j}$ for all $i\in[n]$ are isomorphic. Thus, we can regard each $T_{i,j}$ as a clone of $T_{1,j}$, and provide a simple drawing to express it. In the next section, the example of Fig.~\ref{3CIST-351} will be used again to auxiliary illustrate the construction of multiple CISTs in a high-order RCube. At that time, we will still use this simple drawing for representing CISTs.

\section{Constructing multiple CISTs of RCube with High Order}
\label{sec:main}

\begin{figure*}[ht]
\centering
\includegraphics[width=0.98\textwidth]{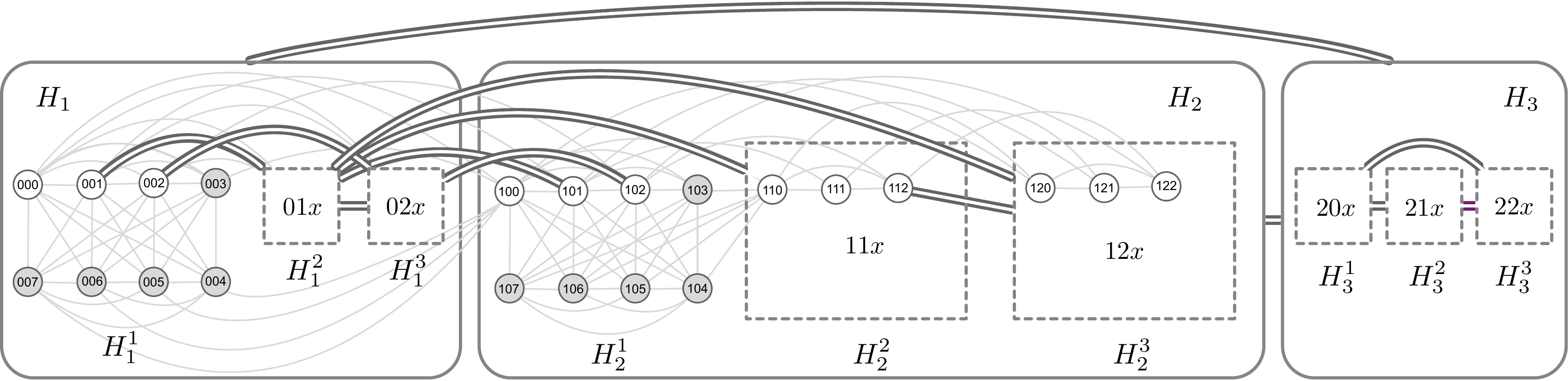}
\caption{A partial view of $L$-$RCube(3,5,2)$, where a round-corner box represents $H_i$ for $i\in[3]$ (i.e., $L$-$RCube(3,5,1)$), a dashed-line box inside $H_i$ represents $H_i^j$ for some $j\in[3]$ (i.e., $L$-$RCube(3,5,0)$), and a parallel line indicates that there exist multiple edges between the two ends of vertices or subgraphs.
}
\label{fig:L-RCube352}
\end{figure*}

In this section, we deal with the construction of CISTs in RCube networks with high-order by recursion. Recall that for $k\geqslant 2$, the logic graph $L$-$RCube(n,m,k)$ is composed of $n$ $L$-$RCube(n,m,k-1)$'s. Similar to the previous section, we use $H_i$ to denote $L$-$RCube(n,m,k-1)$ for $i\in[n]$. Also, each $H_i$ consists of $n^{k-1}$ basic building elements (i.e., $K_{n+m}$), denoted by $H_i^\ell$ for $\ell\in[n^{k-1}]$, in which every vertex has label with the leftmost symbol being $i-1$. Assume $v_{i,j}^\ell\in V(H_i^\ell)$ for $j\in[n+m]$, where we map the index $j\in[n+m]$ of vertices to the servers' addresses in lexicographic order. For instance, Fig.~\ref{fig:L-RCube352} shows a partial view of $L$-$RCube(3,5,2)$, where $H_2$ contains three isomorphic subgraphs $H_2^1$, $H_2^2$ and $H_2^3$, and each subgraph has eight vertices. Similar to Fig.~\ref{3CIST-351}, we use the wildcard symbol $x\in[0,7]$ to denote the rightmost symbol for vertices in a subgraph. In this case, vertices in $H_2^2$ are labeled by $11x$, including three core servers $110$, $111$, and $112$. By Definition~\ref{definition2}(3), the vertex $v_{2,3}^2(=112)$ is adjacent to vertices with labels $12x\in V(H_2^3)$ for $x\in[3,7]$, and thus we draw the multiple edges between $v_{2,3}^2$ and the subgraph $H_2^3$ by a parallel line. Similarly, there exist edges joining vertices with labels $11x\in V(H_2^2)$ for all $x\in[3,7]$ and a vertex $v_{1,2}^2(=011)\in V(H_1^2)$. In this figure, we omit the parallel line for edges between a specific vertex $v$ and a subgraph $H$ if $v\in V(H')$ and there is parallel line between $H$ and $H'$, e.g., vertex $v_{2,2}^3(=121)\in V(H_2)$ and subgraph $H_3^2$.

In what follows, we still make some preliminary observations on the existence of CISTs, and then propose the algorithm for constructing multiple CISTs in $RCube(n,m,k)$ for $k\geqslant 2$.

\begin{itemize}
\item
For $n=1$ and $m\geqslant 3$, $L$-$RCube(1,m,k)$ is a complete graph with $m+1$ vertices. By Lemma~\ref{lem:Kn}, there are $t=\lfloor (m+1)/2 \rfloor$ CISTs.
\item
For $n\geqslant 4$ and $m=0$, $L$-$RCube(n,0,k)$ is the same as $L$-$BCube(n,k)$. By Lemma~\ref{lem:CIST-BCube}, there are $t=\lfloor n/2 \rfloor$ CISTs.
\end{itemize}

For $n+m\geqslant 4$, $n>1$, and $m>0$, we need to construct the desired CISTs by recursion. Let $s=\lfloor (n+m)/2\rfloor$ and $t=\min\{n,s\}$. The base $k=1$ of recursion can be accomplished by  Algorithm~\ref{alg:L-RCube(n,m,1)}. For $k\geqslant 2$, since $H_i\cong L$-$RCube(n,m,k-1)$ for $i\in[n]$, we assume that $t$ CISTs of $H_i$, say $\tilde T_{i,1},\tilde T_{i,2},\dots,\tilde T_{i,t}$, are generated by recursive calls. For each $j\in[t]$, the subgraph $H_i^\ell$ of $H_i$ contains two inner-vertices $v_{i,j}^\ell$ and $v_{i,j+s}^\ell$ in $\tilde T_{i,j}$. Particularly, we choose both $v_{1,j}^1$ and $v_{1,j+s}^1$ as central vertices when $m=0$; and only $v_{1,j}^1$ as central vertex when $m>0$. Then, by a similar approach of Algorithm~\ref{alg:L-RCube(n,m,1)} that adjusts the connecting edges (resp.\ adds the connecting edges) between central vertices and inner-vertices of subgraphs, respectively, for $m=0$ and $m>0$, we can obtain the spanning tree $T_j$. We give all detailed steps of the construction, as shown in Algorithm~\ref{alg:L-RCube(n,m,k)}.

\begin{algorithm}
\small
\caption{\small Constructing multiple CISTs in $RCube(n,m,k)$ for $k\geqslant 2$}
\label{alg:L-RCube(n,m,k)}
\LinesNumbered
\DontPrintSemicolon
\KwIn{A logic graph $L$-$RCube(n,m,k)$ with $n+m\geqslant 4$ and $k\geqslant 2$, which is decomposed into $n$ subgraphs $H_1,H_2,\ldots,H_n$ and each $H_i (\cong\! L$-$RCube(n,m,k\!-\!1))$ has vertices $v_{i,j}^\ell$ for $\ell\in[n^{k-1}]$ and $j\in[n+m]$.}
\KwOut{A set of $t$ CISTs $\{T_1,T_2,\ldots,T_t\}$, where $t=\lfloor(1+m)/2\rfloor$ if $n=1$; and
$t=\min\{n,\lfloor(n+m)/2\rfloor\}$ otherwise.}

$s\leftarrow\lfloor(n+m)/2\rfloor$;\;
\eIf{$n=1$}{
	$t\leftarrow s$;
}
{
	$t\leftarrow \min\{n,s\}$;
}
\eIf{$k=1$ {\rm or} $n=1$}{
Call {\bf Algorithm~\ref{alg:L-RCube(n,m,1)}} to generate $t$ CISTs $T_1,T_2,\ldots,T_t$;
}
{
	\For{$i\leftarrow 1\ {\bf to}\ n$}
	{
		Recursively call {\bf Algorithm~\ref{alg:L-RCube(n,m,k)}} to generate $t$ CISTs
		$\tilde T_{i,1},\tilde T_{i,2},\dots,\tilde T_{i,t}$ of $H_i (\cong L$-$RCube(n,m,k-1))$,
		where $\tilde T_{i,j}$ for $j\in[t]$ contains $v_{i,j}^1$ and $v_{i,j+s}^1$ as
		inner-vertices;
	}
	\For{$j\leftarrow 1\ {\bf to}\ t$}{
		\eIf{$m=0$}{
			$T_j\!\leftarrow\!\!\left(\bigcup\limits_{i=1}^n\tilde{T}_{i,j}\!\right)\cup
			\left(\bigcup\limits_{i=2}^n\!\big\{(v_{1,j}^1,v_{i,j}^1),
			(v_{1,j+s}^1,v_{i,j+s}^1)\big\}\!\!\right)\setminus
			\left(\bigcup\limits_{i=2}^n\big\{(v_{i,j}^1,v_{i,j+s}^1)\big\}\right)$;
		}
		{
			$T_j\leftarrow\left(\bigcup\limits_{i=1}^n\tilde{T}_{i,j}\right)
			\cup\left(\bigcup\limits_{i=2}^n\big\{(v_{1,j}^1,v_{i,j}^1)\big\}\right)$;
		}
	}
	}
\Return $\{T_1,T_2,\ldots,T_t\}$.
\end{algorithm}

\begin{theorem}\label{thm:4-main}
For $n+m\geqslant 4$ and $k\geqslant 2$, Algorithm~\ref{alg:L-RCube(n,m,k)} construct $t$ CISTs of $L$-$RCube(n,m,k)$ in $\mathcal{O}(n^k(n+m))$ time, where
\[
t=\begin{cases}
\lfloor(1+m)/2\rfloor & \text{if $n=1$}; \\
\min\{n,\lfloor(n+m)/2\rfloor\} & \text{otherwise.}
\end{cases}
\]
\end{theorem}
\pf
Even though this is a recursive algorithm, like the previous section's algorithm, its time complexity is still linearly proportional to the number of servers. Thus, it can be run in $\mathcal{O}(n^k(n+m))$ time.

As to correctness, the special case of $n=1$ does not matter with the recursion. In this case, $L$-$RCube(1,m,k)$ is a complete graph with $m+1$ vertices. Thus, the correctness directly follows from Lemma~\ref{lem:CIST-BCube} and the construction of Algorithm~\ref{algo1}. In other cases, the proof is by induction on $k$. By Theorem~\ref{thm:3-main} and the construction of Algorithm~\ref{alg:L-RCube(n,m,1)}, the correctness of the base case $k=1$ is proved. Suppose that the result of this theorem holds for $L$-$RCube(n,m,k-1)$ with $k\geqslant 2$, and let $\tilde T_{i,1},\tilde T_{i,2},\dots,\tilde T_{i,t}$ be CISTs of $H_i$ (i.e., an $L$-$RCube(n,m,k-1)$) generated from the recursion for $i\in[n]$. We now consider $L$-$RCube(n,m,k)$ through the following two cases.

For $m=0$, $H_i$ is isomorphic to $L$-$BCube(n,k-1)$ and $t=\lfloor n/2 \rfloor$. For each $j\in[t]$, we choose the two inner-vertices $v_{i,j}^1$ and $v_{i,j+s}^1$ as central vertices to connect $\tilde T_{1,j}$ and all other $\tilde T_{i,j}$ for $i\in[2,n]$. Then, we remove the edge $(v_{i,j}^1,v_{i,j+s}^1)$ for all $i\in[2,n]$. Thus, we obtain the resulting tree $T_j$ (see Line~14).

For $m>0$, we have $t=\min\{n,\lfloor(n+m)/2\rfloor\}$. For each $j\in[t]$, we choose $v_{i,j}^1$ as the central vertex to connect $\tilde T_{1,j}$ and all other $\tilde T_{i,j}$ for $i\in[2,n]$ through the edges $(v_{1,j}^1,v_{i,j}^1)$. Consequently, we obtain the resulting tree $T_j$ (see Line~16).

By the induction hypothesis, $\tilde T_{1,j},\tilde T_{2,j},\dots,\tilde T_{n,j}$ are edge-disjoint in $H_i$ for $i\in[n]$. In the above two cases, since all adjusting edges or connecting edges between trees are different, this implies that $T_1,T_2,\ldots,T_t$ are edge-disjoint spanning trees. Moreover, the construction of $T_j$ for $j\in[t]$ merely joins edge with two inner-vertices, no leaf becomes to inner-vertex in this process. By Theorem~\ref{thm:Hasunuma}, it guarantees that $T_1,T_2,\ldots,T_t$ are CISTs of $L$-$RCube(n,m,1)$.
\qed

Taking the result of Remark~\ref{rmk:diameter} as a base, since the increment of diameter of a spanning tree in each recursion is 2, we obtain the following result.

\begin{remark}\label{rmk:diameter-k}{\rm
For $L$-$RCube(n,m,k)$ with $n+m\geqslant 4$, let $T$ be a spanning tree constructed by Algorithm~\ref{alg:L-RCube(n,m,k)}. Then, $T$ has the diameter as follows:
\[
\text{diam}(T)=\begin{cases}
3 & \text{if $n=1$}; \\
2k+3 & \text{if $n=2$ or $m=0$}; \\
2k+4 & \text{otherwise.}
\end{cases}
\]
}
\end{remark}

\begin{figure*}[ht]
\centering
\includegraphics[width=0.9\textwidth]{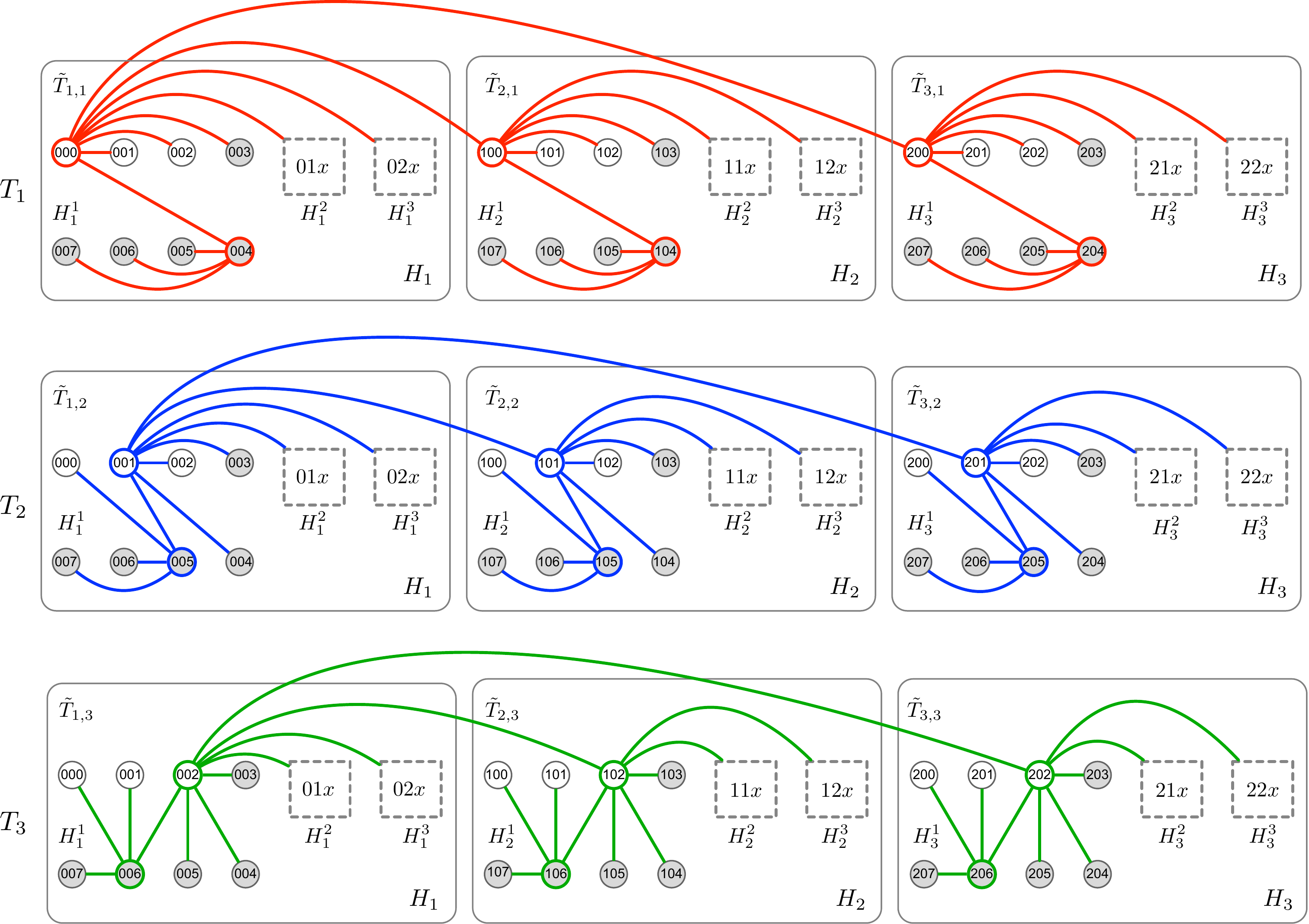}
\caption{Three CISTs of $L$-$RCube(3,5,2)$ constructed by Algorithm~\ref{alg:L-RCube(n,m,k)}.}
\label{fig:352}
\end{figure*}

We close this section by providing a complete example of CISTs for $L$-$RCube(3,5,2)$, as shown in Fig.~\ref{fig:352}.

\section{Application of Multi-Protection Routing and Its Performance Evaluations}
\label{sec:App}

In this section, we experimentally assess multi-protection routing performance using simulation results for three types of transmission. As to the choice of evaluating performance, we are interested in computing the \emph{transmission failure rate} (TFR for short) and its related extension topics. TFR is the ratio of the number of failed transmissions to the overall communications and is one of the essential factors for network reliability and quality-of-service (QoS). As mentioned earlier, there are three types of transmission in the environment of a DCN with heterogeneous edge-core servers. By mapping various applications to the three types of transmission (e.g., see the description in Fig.~\ref{fig:edge-core}), we observe that Type-1 apps (resp.\ Type-2 apps) use only edge-edge transmission (resp.\ edge-core transmission). However, Type-3 apps mixed-use edge-core and core-core transmissions. To make the considerations more realistic in simulation, we design three models with a different transmission amount for different transmission types according to the trigger events. We assume that reaching half of the transmission amount is heavy usage, the rate of 35\% is normal usage, and not over 15\% is light usage. From experience, although a large number of edge-edge transmissions occur, each of these transmissions has a small conveying quantity, and thus the total amount is not greater than the other two types. Table~\ref{tbl:3model} lists all ratios of the three scenarios models for simulation.

\begin{table}[ht]
\caption{The ratios of trigger events for the different transmission types in three scenarios models of simulation: Heavy usage 50\%, Normal usage 35\%, and Light usage 15\%.}
\begin{center}
\begin{tabular}{|c|ccc|c|}
\hline
Transmission types & \!Model~1\! & \!Model~2\! & \!Model~3\! & For Apps \\
\hline
Edge-Edge & 15\% & 15\% & 30\% & Type-1 \\
Edge-Core & 35\% & 50\% & 35\% & Type-2 \& Type-3 \\
Core-Core & 50\% & 35\% & 35\% & Type-3  \\
\hline
\end{tabular}
\end{center}
\label{tbl:3model}
\end{table}%

Technically, all algorithms for constructing CISTs and the required routing algorithms are implemented by using Python programs. We carry out the simulation by using a 2.00GHz $\text{Intel}^{\circledR} \text{Xeon}^{\circledR}$ Gold 5117 CPU and 32 GB RAM 
under the Linux operating system. To design a multi-protection routing, Pai et al.~\cite{Pai-2020-JPDC} extended Tapolcai's method (introduced in Section~\ref{sec:protection-route}). They called such an extension the \emph{multi-protection routing scheme} (MPR-scheme for short), which is configured by a combination of multiple CISTs and can be used to increase fault-tolerance capability. We follow the use of MPR-scheme in our simulation.

For each $L$-$RCube(n,m,k)$ with $n,m\in[3,5]$ and $k,\ell\in[3]$ (where $\ell$ indicates the scenarios model listed in Table~\ref{tbl:3model}), we first construct $t=\min\{n,\lfloor(n+m)/2\rfloor\}$ CISTs by using Algorithms~\ref{alg:L-RCube(n,m,1)} and \ref{alg:L-RCube(n,m,k)}, determining by $k=1$ or $k\geqslant 2$. For each situation, we randomly generate 100,000 instances of vertex pairs $(s,d)$ with $s\ne d$, where $s$ and $d$ are the source and the destination of transmission, respectively. Hereafter, we always assume that $s$ and $d$ are fault-free. Moreover, $s$ and $d$ to be edge servers or core servers are chosen by the model's ratio in Table~\ref{tbl:3model}, according to $\ell$. Besides, we randomly choose a set $F$ of faulty vertices to inspect whether the failure occurs in the transmission from $s$ to $d$. The vertices $s$ and $d$ (resp.\ vertices in $F$) are uniformly distributed over the chosen servers (resp.\ the whole network). Let $r={t\choose 2}$. Since we can configure a protection routing for any two of the CISTs using Algorithm~\ref{Tapolcai}, we have totally $r$ distinct protection routings when we adopt MPR-scheme. For $i\in[r]$, let $R_i$ be the configured protection routings and $P_i$ be the transmitting path from $s$ to $d$ in $R_i$. A transmitting path is called a \emph{regular route} provided $V(P_i)\cap F=\emptyset$, and a \emph{failed route} otherwise. Note that, for a fault-tolerant routing, we may perform the transmission from $s$ to $d$ in $P_i$, simultaneously, for all $i\in[r]$. Thus, if there exist $r-1$ failed routes in the transmission, we can still acquire a successful transmission. Consequently, it guarantees that the MPR-scheme can tolerate $t-1$ faulty vertices in the transmission. For each $RCube(n,m,k)$ with different $n,m$, and $k$, we start the simulation from two-node failures and keep on adding one node failure at one time in $F$ until failures reach at least ten nodes and TFR is over a threshold, which we set up 0.5\%, for all three models.

Due to the space limitation, the complete simulation results of multi-protection routings for $L$-$RCube(n,m,k)$ are available on the website~\cite{Li-2021-web}. Here we only extract parts of the required results to conduct the analysis. Firstly, the intuitive idea is that TFR should be presented as an increasing function with respect to the number of faulty nodes. As expected, our simulations are almost consistent. For example, Fig.~\ref{fig:Sim-1} shows the cases of $L$-$RCube(5,m,3)$ for $m\in[3,5]$, and refer to~\cite{Li-2021-web} for complete cases. In actual simulation, TFR does not necessarily increase with the number of faulty nodes. For instance, see Fig.~\ref{fig:Sim-1}(a) for $L$-$RCube(5,3,3)$. We have $\text{TFR}=0.00561$ when $|F|=70$ and $\text{TFR}=0.00497$ when $|F|=71$ in the model~3. Also, see Fig.~\ref{fig:Sim-1}(c) for $L$-$RCube(5,5,3)$. We have 481 transmission failures when we deal with $|F|=121$ in model~1. Thus, we can compute $\text{TFR}=481/10^5=0.00481$. Moreover, we have the following analyses:

\begin{figure}[h]
\centering
\includegraphics[width=0.95\textwidth]{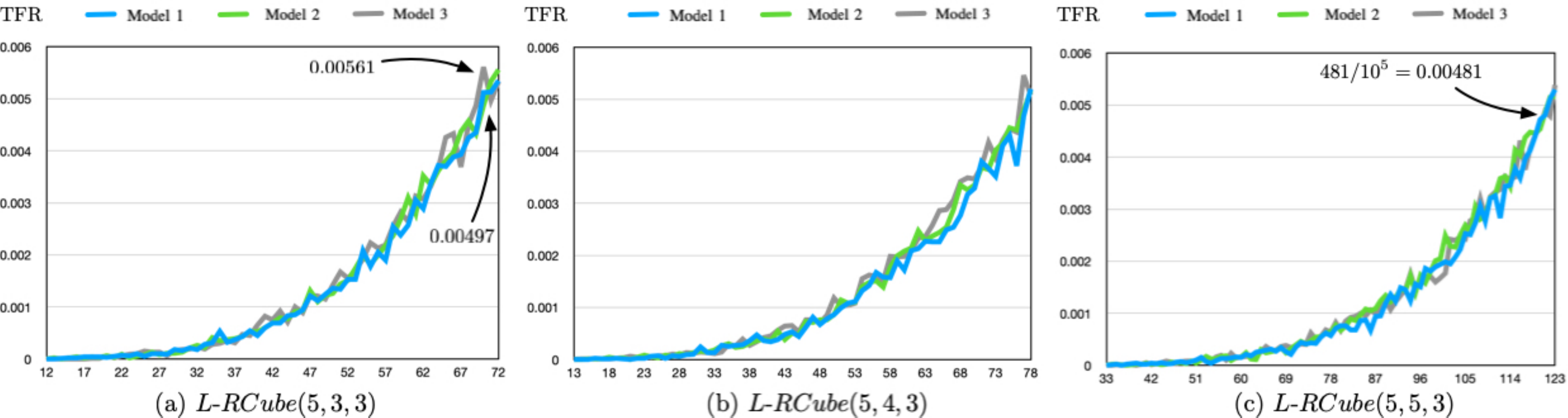}
\caption{Variation of TFR under different numbers of faulty nodes in $L$-$RCube(5,m,3)$ for $m=\{3,4,5\}$.}
\label{fig:Sim-1}
\end{figure}

\begin{figure}[h]
\centering
\includegraphics[width=0.95\textwidth]{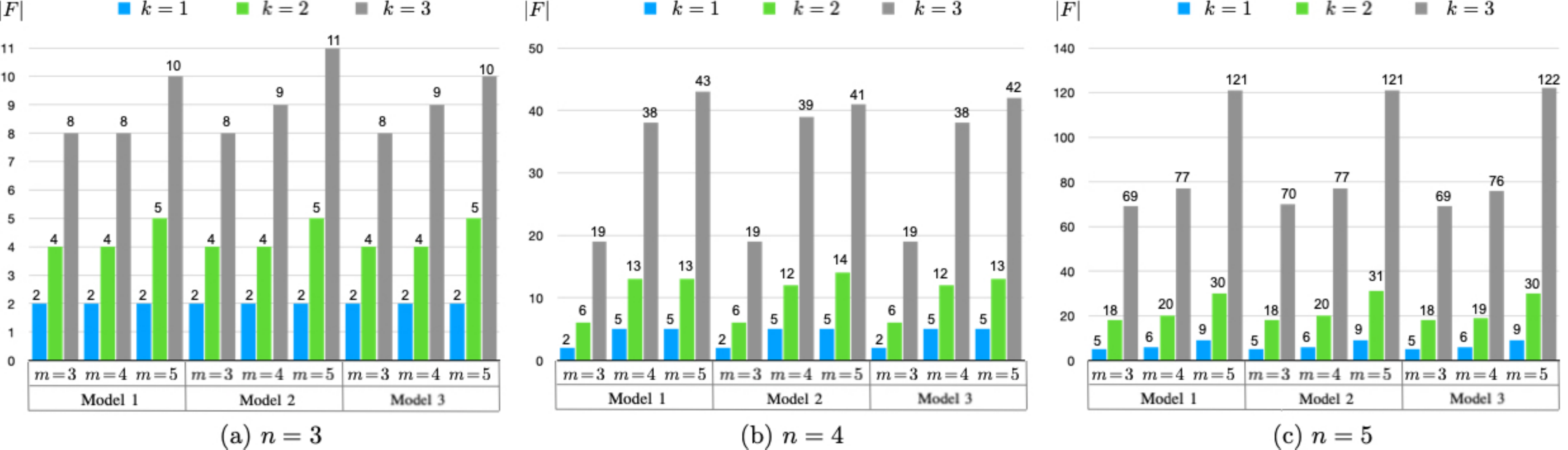}
\caption{The maximum number of faulty nodes accommodated in the three models of $L$-$RCube(n,m,k)$ when the threshold does not exceed 0.5\%.}
\label{fig:Sim-2}
\end{figure}

(i) For observing the variation of the fault-tolerant effect for the three parameters $n,m$, and $k$ in $L$-$RCube(n,m,k)$, we show the maximum number of faulty nodes accommodated in the three models when its threshold does not exceed 0.5\% in Fig.~\ref{fig:Sim-2}. For instance, we consider $n=m=5$ in model~1 (see Fig.~\ref{fig:Sim-2}(c)). Then, the multi-protection routing can tolerate $9,30$, and $121$ faulty nodes for $k=1,2,3$, respectively. Also, if we consider $n=5$ and $k=3$, then the multi-protection routing can tolerate $69,77$, and $121$ faulty nodes in the same model for $m=3,4,5$, respectively. Finally, if we consider $m=5$ and $k=3$, then the multi-protection routing can tolerate $10,43$, and $121$ faulty nodes in model~1 for $n=3,4,5$, respectively. Therefore, we aware of the following phenomenon: for the three parameters $n,m,k$, if two of them are fixed, then the multi-protection routing with the higher value of the third parameter can tolerate more faulty nodes. This shows the fact that the number of faulty nodes accommodated is positively correlated to the network scale. Since the total number of nodes in $L$-$RCube(n,m,k)$ is $N=(n+m)n^k$, the change of $k$ has the most influence on $N$, followed by $n$, and $m$ has the less impact. Obviously, the changes in the above three sequences $(9,30,121)$, $(10,43,121)$ and $(69,77,121)$ actually present this appearance.

\begin{figure}[h]
\centering
\includegraphics[width=0.95\textwidth]{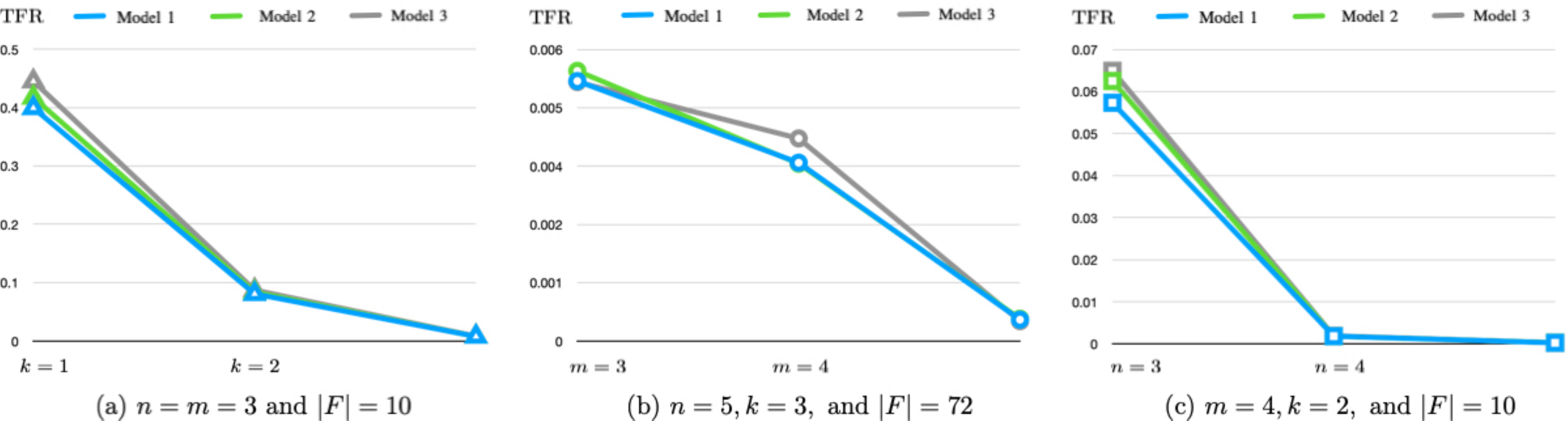}
\caption{TFR presents a negative correlation with the third parameter when the other two parameters and the number of faulty nodes are fixed.}
\label{fig:Sim-3}
\end{figure}

(ii) As observed before, TFR is almost positively related to the number of faulty nodes. From (i), we may be anxious that TFR is also positively correlated with the network scale. In fact, it is the opposite. When the number of faulty nodes is fixed, the average density of faulty nodes on a network with a larger scale will decrease, so the TFR will also be lower. We examine the tendency of TFR when only one parameter of $n,m,k$ is altered in the three models of $L$-$RCube(n,m,k)$. Fig.~\ref{fig:Sim-3}(a) shows the case of $n=m=3$ and $|F|=10$; Fig.~\ref{fig:Sim-3}(b) is the case of $n=5$, $k=3$, and $|F|=72$; and Fig.~\ref{fig:Sim-3}(c) is the case of $m=4$, $k=2$, and $|F|=10$. Only three cases have been cited here, but in reality, all cases are similar. For the three parameters $n,m,k$, if two of them are fixed under the same number of faulty nodes, then the higher value of the third parameter possesses the lower TFR. Therefore, we conclude that if the same number of nodes failed, the larger the network scale, the smaller the TFR.

\begin{figure}[h]
\centering
\includegraphics[width=0.95\textwidth]{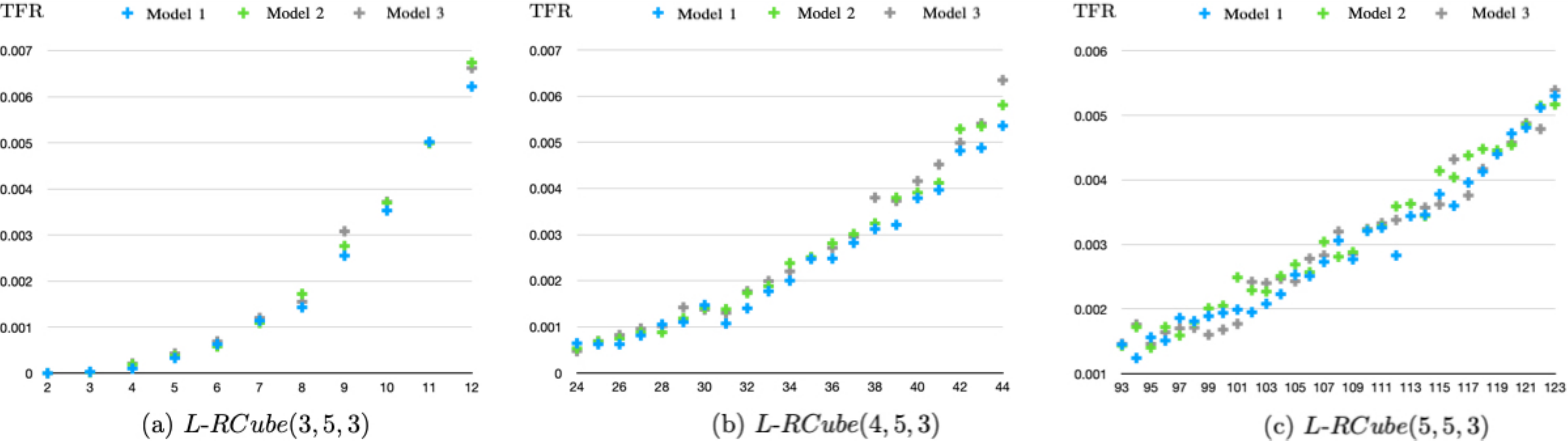}
\caption{Variation of TFR under different models in $L$-$RCube(n,5,3)$ for $n=\{3,4,5\}$.}
\label{fig:Sim-4}
\end{figure}

(iii) Finally, we will focus on analyzing TFR in different models of $L$-$RCube(n,m,k)$, which is the matter of our core concern. As before, we only appoint three cases here, but actually, all cases are similar. Fig.~\ref{fig:Sim-4}(a)-(c) show the variation of TFR under the three models in $L$-$RCube(n,5,3)$ for $n=3,4,5$, respectively. The samples we extracted have a range of $10(n-2)$ faulty nodes before the threshold being reached. From the simulation results, we may observe that the TFR of Model~1 is almost the lowest in most cases, while the other two are indistinguishable. In the above-constructed protection routings $R_i$ ($i\in[r]$), core servers are generally connected with multiple servers and play a relay role in the communication of $R_i$. In contrast, the edge servers have fewer connections in $R_i$. Thus, if there are failures and most of them are core servers, it will cause a severe impact on the data transmission, whereas it will have a minor effect if most of them are the edge servers. Recall that $s$ and $d$ to be core servers or edge servers are chosen by different models in the simulation. Since the proportion of core-core in Model~1 is relatively large and $s$, $d$, and all faulty nodes are generated by the uniform distribution, this causes that TFR is relatively small. Also, the proportion of edge-core in Model~2 (resp.\ edge-edge in Model~3) is relatively large compared to other models, and thus the result will be the opposite. For this phenomenon, we have the following results in most cases under the assumption in Table~\ref{tbl:3model}:
TFR(Model~1) $<$ TFR(Model~2) and TFR(Model~1) $<$ TFR(Model~3).

\section{Concluding remarks}
\label{sec:Conclusion}
First of all, we point out that the heterogeneous servers configured by RCube are suitable as a candidate topology of DCNs for edge computing to serve different kinds of applications. Then, we propose a construction of multiple CISTs in $RCube(n,m,k)$ for $n+m\geqslant 4$, and discuss the diameters of CISTs we constructed. As a by-product, such a construction can provide a feasible protection routing of the network. Moreover, we extend Tapolcai's method and show that a protection routing configured by a combination of multiple CISTs (which was called MPR-scheme) can further increase network communication capability. Most importantly, it can improve the fault-tolerance of transmission. Theoretically, the MPR-scheme configured by $t$ CISTs can tolerate $t-1$ failed components, so no failed transmission will occur. From the simulation experiments, the routing using MPR-scheme indeed achieves better fault-tolerance efficiency (e.g., for $L$-$RCube(5,5,3)$, it contains 1250 servers and can tolerate up to 121 failed nodes such that TFR in the three models does not exceed 0.5\%). In particular, we have the following results for the multi-protection routing in our simulations:

\begin{itemize}
\item The number of faulty nodes accommodated is positively correlated to the network scale.
\item The larger the network scale, the smaller the TFR (when the number of faulty nodes is fixed).
\item When the heavy usage of core-core transmission reaches half of the transmission amount, it still maintains the lowest TFR in most cases.
\end{itemize}

Finally, we conclude this paper by discussing some drawbacks to the multi-protection routing. Inevitably, the adoption of the MPR-scheme requires a large amount of computation time. What's worse, we have found that the entire network will be full of packets in our experiment and lead to heavy traffic. Therefore, how to reduce network traffic under the multi-protection routing is a problem needed to address in the future. Can we add a congestion control mechanism for which a preset diversion condition will split the stream of packets and continue to deliver until all packets reach their destinations to achieve load balancing?

\section*{Acknowledgments}
This work was supported by the National Natural Science Foundation of China (Nos. 61872257, 62002062 and 62072109) and the Ministry of Science and Technology of Taiwan (No. MOST-107-2221-E-141-001-MY3).

\end{document}